# *Fresnel Drag in the Homogenization Limit with Space-Time-Modulated Wire Media*

Alexander B. Yakovlev[1,*] , Mário G. Silveirinha[2], Arghyadeep Pal[1], and Mehdi H. Zadeh[1]

[1]Department of Electrical and Computer Engineering, University of Mississippi, University, Mississippi, 38677-1848, USA

[2]University of Lisbon – Instituto Superior Técnico and Instituto de Telecomunicações, Avenida Rovisco Pais 1, 1049-001 Lisbon, Portugal

## Abstract

Space–time modulations of the electromagnetic response offer new opportunities for wave control. In particular, such systems can emulate moving-medium responses and the associated Fresnel drag in the homogenization limit. Existing approaches require the simultaneous microscopic modulation of both permittivity and permeability, which is difficult to realize in practice. Here, we show that modulating a metallic response overcomes this limitation and enables strong moving-medium-like effects using purely electric modulation. We illustrate this mechanism with a space–time-modulated wire medium, described through Lorentz transformations and quasi-static homogenization. The resulting effective medium is nonreciprocal and bianisotropic and supports a pronounced synthetic Fresnel-drag effect. For a finite-thickness slab, this response leads to nonreciprocal scattering while preserving global energy conservation for propagating waves. Remarkably, the synthetic Fresnel drag also produces velocity-dependent reflection and transmission Goos–Hänchen shifts, providing a direct signature of the effective motion.

* Corresponding author: yakovlev@olemiss.edu

# I. Introduction

With the rapid pace of advancements in electromagnetic materials research, time-varying and space-time-modulated metamaterials and metasurfaces have received considerable attention [1-6]. Space-time-varying metamaterials/metasurfaces enable unprecedented wave propagation control beyond the limits of passive, time-invariant structures, emerging as "four-dimensional" (4D) metamaterials engineering [1,3,7,8]. The essence of space-time-modulated media is that it breaks classical Lorentz reciprocity theorem providing additional degrees of freedom for design of novel magnet-less nonreciprocal devices [7]. 4D metamaterial crystals are described through dispersion relations that live in a higher-dimensional Brillouin zone, where energy and momentum take generalized forms. This framework enables engineering nonreciprocity, gain responses, and even topological phase transitions in frequency-momentum space [4,9].

Building on these concepts, researchers have developed effective-medium and homogenization theories in 4D space-time frameworks to describe metamaterials under spatio-temporal modulation.  A space-time-modulated medium can behave as if it were physically moving at an effective velocity set by the modulation pattern. This produces Doppler-like frequency shifts and a Fresnel-Fizeau synthetic drag (nonreciprocal phase accumulation) for waves propagating along the direction parallel to the modulation velocity [7,10,11].

A useful approach is to perform a Lorentz (or Galilean) transformation to a co-moving frame where the modulation is static, allowing one to map a space-time-periodic to an equivalent bianisotropic medium at rest [7]. A homogenized description of travelling-wave modulations that yields effective permittivity, permeability, and magnetoelectric coupling tensors dependent on the modulation velocity has been derived in [7], revealing that a travelling refractive-index grating induces giant bianisotropy (magnetoelectric coupling) as a direct signature of synthetic motion, and extreme Fresnel drag even for weak modulations. Also, in [12,13] isorefractive spacetime crystals have been proposed as travelling-wave modulated lattices that exactly mimic the electrodynamics of a physically moving medium as seen by a laboratory observer. This establishes a formal moving-medium analogue: a modulated medium can reproduce all optical effects of moving matter (such as aberration and Fresnel-Fizeau drag) without actual bulk motion. Such insights underscore the potential of space-time metamaterials as a platform to study relativistic phenomena (including relativistic Doppler effects, synthetic event horizons, Hawking-like radiation) within a stationary structure [7]. More broadly, the homogenization of time-variant and space-time variant systems has been discussed previously by different groups [7,10,12-20].

A key limitation of the approaches developed so far to realizing moving-medium analogues with space–time crystals is that, they require the explicit and simultaneous modulation of both the microscopic permittivity and permeability to reproduce a moving-medium response in the homogenization limit [7,11-13]. This requirement is difficult to meet experimentally because independently controlling electric and magnetic responses is generally challenging. Here, we show that this limitation can be overcome by modulating a metallic-like response. Using a space–time-modulated wire medium as a representative platform, we demonstrate that modulation of the electric response alone is sufficient to generate an effective nonreciprocal and bianisotropic medium with strong magnetoelectric coupling. In principle, such moving-like spatiotemporal responses may be engineered at microwave frequencies by loading metallic wires with switches that are periodically turned on and off in a synchronized manner. Related experimental realizations of time-modulated media at microwave frequencies have been demonstrated using switched transmission-line and microstrip platforms [21-23]. In our proposal, the homogenized medium exhibits a pronounced synthetic Fresnel-drag effect, which originates nonreciprocal scattering and velocity-dependent Goos–Hänchen shifts. Our approach provides a practical route to moving-medium analogues based entirely on modulated metallic structures.

The wire medium is an artificial material composed of 2D lattice of parallel thin metallic wires (see Fig. 1a), and has been of interest for a long time [24,25]. In the last two decades it has gained attention due to advancements in metamaterials research at microwave, terahertz, and optical frequencies with a variety of applications, including negative refraction [26-29], ε-near-zero materials [30,31], subwavelength imaging of the near field [32-35], canalization and transport of the near field [36-38], high-impedance mushroom substrates [39-41], broadband microwave absorbers [42,43], gap waveguides [44-46], among others. A comprehensive review of wire-medium metamaterials at terahertz and optical frequencies is given in [47]. With subwavelength dimensions (radius and period of wires) uniaxial wire media can be homogenized as an anisotropic material characterized by a spatially dispersive effective permittivity function along the wires. In particular, at microwave frequencies the wire medium is characterized by strong spatial dispersion even for very long wavelengths [48,49]. Previous work developed non-local homogenization models for the accurate analysis of the interaction of electromagnetic waves with wire media and for solving scattering, radiation, and excitation problems [27-29,32-35,40,41,43,50-57]. Also, for bounded wire-medium structures, the non-local homogenization formalism requires the additional boundary conditions at material interfaces to account for the extra waves arising from nonlocality [58-61].

Here, we investigate the effect of space–time modulation on the electromagnetic response of a wire medium from a homogenization perspective. The formulation is based on the quasi-static homogenization model for wire media [51,52,60], with the governing equations for fields, current, and additional potential derived in the laboratory and co-moving frames with the use of Lorentz transformations. In our analysis we exploit the fact that the perfect electric conductor (PEC) boundary condition is Lorentz invariant, i.e., frame independent, when the velocity is parallel to the boundary [17]. We demonstrate that because of this equivalence a space-time modulated wire medium can mimic exactly a physically moving medium, analogous to isorefractive crystals [12] and to the space-time diffraction gratings studied in [17]. In our analysis, we assume that the host material is air and model the wires as thin metallic strips parallel to the direction of the motion (see Fig. 1c), rather than using the more conventional cylindrical geometry (Fig. 1a). This is motivated by the fact that both air and the PEC strips have an electromagnetic response that is Lorentz invariant, i.e. it is identical in the laboratory (Fig. 1c) and the co-moving frames (Fig. 1b).

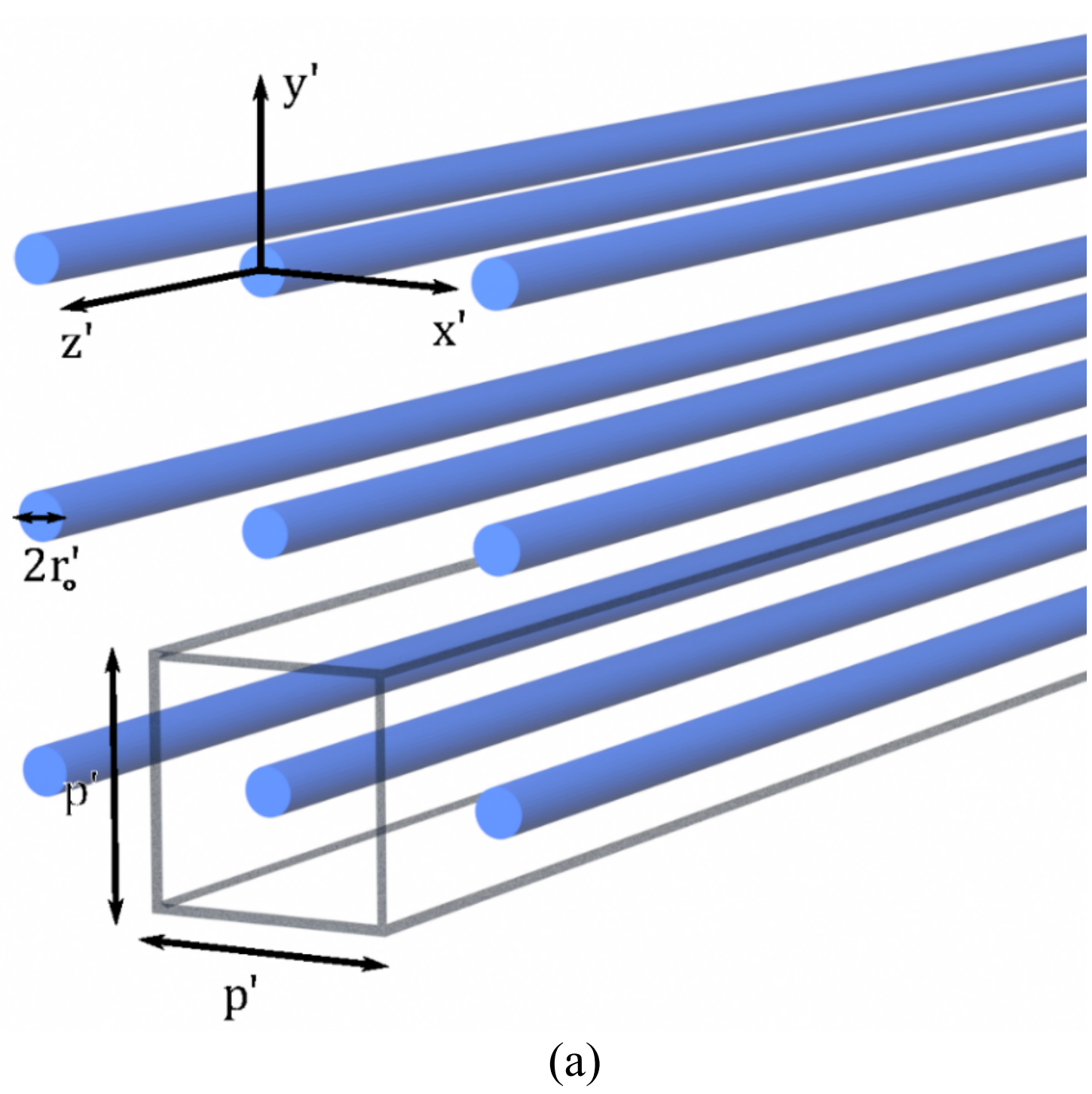

(a)

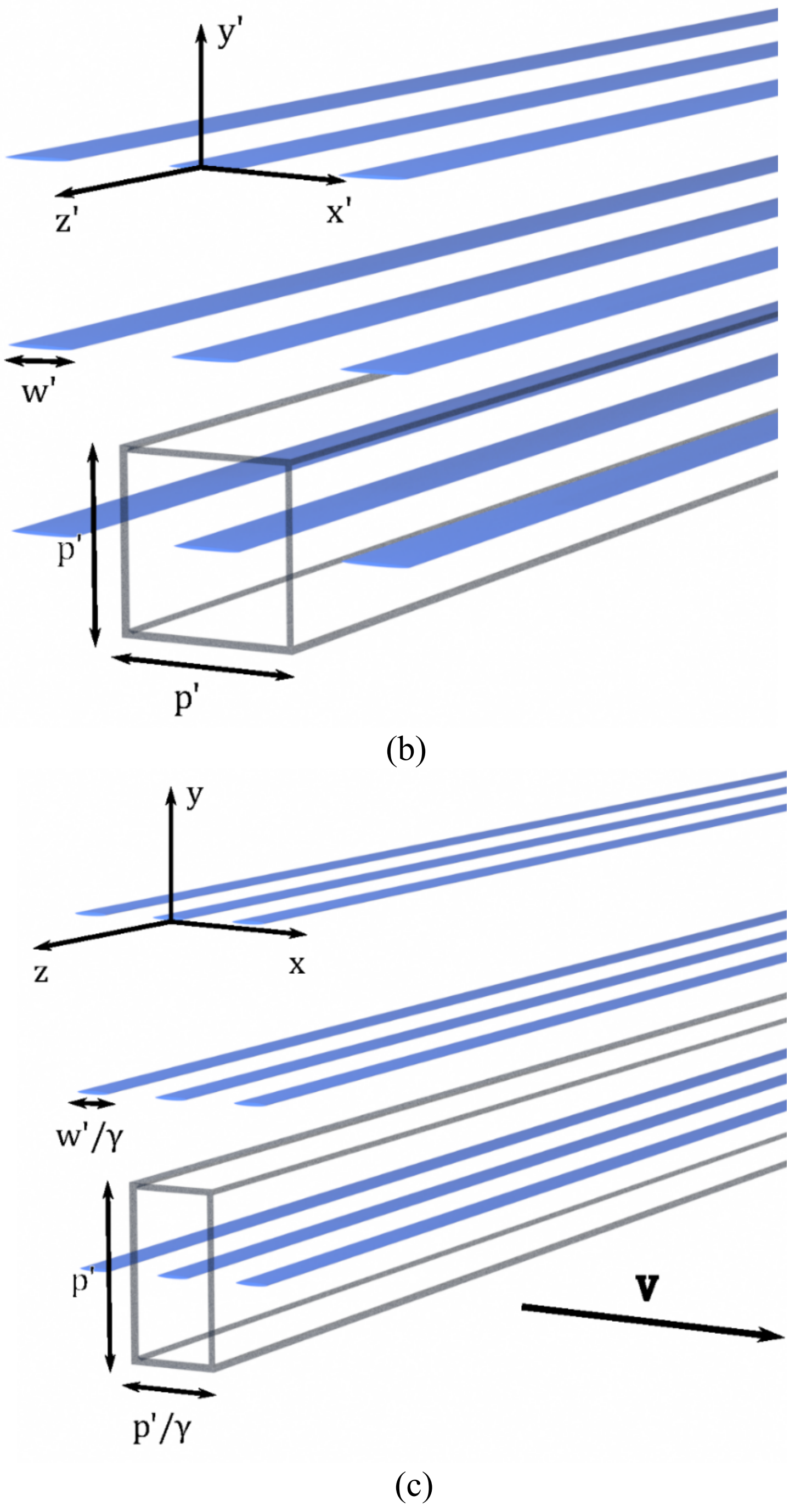

**Fig. 1** (a) Geometry of uniaxial wire media formed by a 2D lattice of parallel thin metallic wires in the static (co-moving) frame; (b) Geometry of the equivalent wire media in the quasi-static approximation formed by thin metallic strips along the direction of the motion with $w' = 4r_0'$ in the co-moving frame; (c) Geometry of the equivalent wire media in the laboratory frame moving with the velocity $\mathbf{v} = \hat{\mathbf{x}}v$. Due to Lorentz-Fitzgerald contraction [62], the strip width $w'$ and the period $p'$ in the $x-$direction are contracted resulting in the rectangular lattice in the laboratory frame, such that $w = w'/\gamma$, $p = p'/\gamma$. Here, $\gamma$ is the Lorentz factor, $\gamma = \left(1 - v^2/c^2\right)^{-1/2}$.

Our homogenization theory shows that, in the quasi-static limit, the space–time-modulated wire medium exhibits an effective nonreciprocal and bianisotropic response. As in conventional static wire media, this response remains strongly spatially dispersive and therefore cannot, in general, be described by a local constitutive model. More broadly, nonlocality or spatial dispersion play an important role in space-time-modulated metamaterials and metasurfaces [14-16]. This can be understood by analogy with moving systems, where frequency dispersion in the co-moving frame necessarily leads to a nonlocal response in any other inertial frame. In space–time-modulated wire media, these effects are further enhanced by the intrinsic nonlocality of the underlying wire-medium response.

Related spatiotemporally modulated wire media were first investigated in Ref. [10] using a Floquet–Bloch formulation for an array of capacitively loaded wires. Mode-dependent effective permittivities were extracted from the corresponding dispersion branches. However, the analysis of Ref. [10] is restricted to propagation in the plane perpendicular to the wires. Reference [10] also reported an extreme Fresnel-drag effect for in-plane propagation, manifested as a nonreciprocal phase shift even for slow modulations. In contrast, our work develops a constitutive homogenization description of the metamaterial, with particular emphasis on the extraordinary mode that propagates along the direction of the wires. Rather than extracting effective parameters separately from individual dispersion branches, we derive a single nonlocal material matrix in the laboratory frame that describes all supported modes within the same constitutive framework. This formulation further shows that, in the quasi-static limit, the space–time-modulated wire medium is electromagnetically equivalent to an actual moving wire medium. Thus, while the occurrence of Fresnel drag in modulated wire media was established in Ref. [10] for transverse propagation, our theory identifies the moving-medium origin of the drag effect experienced by the axial extraordinary mode and provides a global homogenized description that is not tied to any individual dispersion branch. It should be noted that the wire medium provides a specific realization of the general principle that moving-medium-like responses can be engineered by modulating metallic-like systems, without requiring the simultaneous modulation of permittivity and permeability. Its large intrinsic anisotropy potentially enhances the resulting Fresnel-drag effect.

Using our formalism, we derive the energy conservation laws, demonstrating that the energy balance in the laboratory frame may generally be disrupted in the presence of the synthetic motion. We study the scattering of electromagnetic waves from a space-time finite-thickness wire-medium slab and demonstrate nonreciprocal reflection and transmission in the laboratory frame. Interestingly, notwithstanding the non-Hermitian nature of the system, we show that for plane wave illumination there is no net energy generated or absorbed by the "moving" wire medium. We demonstrate that the reflection and transmission Goos-Hänchen shifts in the reflected and transmitted fields can be controlled by the modulation velocity, and provide a direct fingerprint of the synthetic Fresnel drag.

## II. Quasi-Static Model for Space-Time-Modulated Wire Media

Here, we develop a formalism for the analysis of electromagnetic response of homogenized spatiotemporal uniaxial wire media with the geometry in the static (co-moving) and laboratory frames shown in Fig. 1. In the analysis of spatiotemporal wire media considered in this paper, the conventional cylindrical wires are replaced by equivalent strips (Fig. 1(b)) oriented parallel to the direction of the motion with the velocity $\mathbf{v} = \hat{\mathbf{x}}v$ (Fig. 1(c)). Under these conditions, both the air region and the metallic strips (taken as perfect conductors) have responses that are frame independent, and hence the space-time modulated system is strictly equivalent to a moving system [12,17]. Thereby, the system analysis can be conveniently done in a Lorentz co-moving frame [12]. Owing to the relativistic nature of the transformation, the laboratory-frame and co-moving-frame geometries are related by Lorentz contraction along the direction of motion [12]. For simplicity, we assume that the metallic strips form a square lattice in the co-moving frame. The Lorentz contraction then maps this arrangement onto a rectangular lattice in the laboratory frame.

Our formulation is based on the quasi-static model for wire media [51,52,60] such that the macroscopic electromagnetic fields in the co-moving (primed) frame for the case of wires oriented along the $z-$direction satisfy the system of curl equations:

$$\nabla' \times \mathbf{E}' = -\partial_{t'} \mathbf{B}',$$

(1)

$$\nabla' \times \mathbf{B}' = \frac{1}{c^2} \partial_{t'} \mathbf{E}' + \mu_0 \frac{I'}{A'_c} \hat{\mathbf{z}}. \tag{2}$$

In order to take into account spatial dispersion of wire media, two additional transmission-line equations are introduced for the current $I'$ and the additional potential $\varphi'_w$:

$$\partial_{z'} \varphi'_w = -L' \partial_{t'} I' + E'_z, \tag{3}$$

$$\partial_{z'} I' = -C' \partial_{t'} \varphi'_w. \tag{4}$$

Here, $A'_c = p'^2$ is the area of the unit cell in the co-moving frame, $C' = 2\pi\varepsilon_0 / \log\left[p'^2 / 4r'_0\left(p' - r'_0\right)\right]$ and $L' = \left(\mu_0 / 2\pi\right)\log\left[p'^2 / 4r'_0\left(p' - r'_0\right)\right]$ are the effective capacitance and the effective inductance, respectively, per unit length of a wire [51]. Here, $r'_0$ represents the equivalent radius of the wires, which is related to the strip width by $w' = 4r'_0$. It is underlined that the above set of equations already provides the homogenized response of the metamaterial in the co-moving frame coordinates.

In order to characterize the homogenized response in the laboratory response we use a Lorentz transformation. Under a Lorentz boost the space-time coordinates in the co-moving (primed) frame are related to the space-time coordinates in the laboratory (unprimed) frame by the set of equations [62,63]: $x' = \gamma\left(x - vt\right)$, $y' = y$, $z' = z$, $t' = \gamma\left(t - xv/c^2\right)$, such that in the laboratory frame Eqs. (1)-(2) become

$$\nabla \times \mathbf{E} = -\partial_t \mathbf{B}, \tag{5}$$

$$\nabla \times \mathbf{B} = \frac{1}{c^2}\partial_t \mathbf{E} + \mu_0 \left( \frac{I'}{A'_c}\hat{\mathbf{z}} + \gamma v \rho' \hat{\mathbf{x}} \right) \tag{6}$$

where Eqs. (1) and (5) are Lorentz-invariant and Eq. (6) has a current component along the $x-$ direction. This additional component of the current can be attributed to the effective motion of the electric charges on the metallic strips due to the space-time modulation. The electromagnetic fields transform in the usual way under the Lorentz transformation [62,63]. We took into account that the 4-current ($\left(c\rho, \mathbf{j}\right)$) transforms as a 4-vector and that $\mathbf{j}' = \frac{I'}{A'_c}\hat{\mathbf{z}}$. In particular, the current density along $z$ is unaffected by the transformation.

Furthermore, the continuity equation in the co-moving frame with the current along the wires in the $z-$direction can be written as $\partial_{t'}\rho' + \nabla \cdot \mathbf{j}' = 0 \rightarrow \partial_{t'}\rho' + \frac{1}{A'_c}\partial_{z'} I' = 0$. Using Eq. (4) one obtains a relation between the charge density $\rho'$ and the additional potential $\varphi'_w$: $\partial_{t'}\rho' - \frac{1}{A'_c}C'\partial_{t'}\varphi'_w = 0$, i.e. $\rho' = \frac{1}{A'_c}C'\varphi'_w$. Hence, the system of Eqs. (5)-(6) can be rewritten as follows,

$$\nabla \times \mathbf{E} = -\partial_t \mathbf{B}, \tag{7}$$

$$\nabla \times \mathbf{B} = \frac{1}{c^2}\partial_t \mathbf{E} + \mu_0 \frac{1}{A_c'}\left( I'\hat{\mathbf{z}} + \gamma v C' \varphi_w' \hat{\mathbf{x}} \right). \tag{8}$$

It can be seen that the additional current component, due to the "motion" of the charges on the strips, is dependent on the additional potential.

On the other hand, as the 4-gradient $\left(-\frac{1}{c}\partial_t, \nabla\right)$ also transforms as a 4-vector, we have $\partial_{z'} = \partial_z$, $\partial_{t'} = \gamma\left(\partial_t + v\partial_x\right)$. Taking also into account that $E_z' = \gamma\left(E_z + vB_y\right)$ we obtain two additional equations for the current and the additional potential in the laboratory frame:

$$\partial_z \varphi_w' + \gamma L' v \partial_x I' = -\gamma L' \partial_t I' + \gamma\left(E_z + vB_y\right), \tag{9}$$

$$\partial_z I' + \gamma C' v \partial_x \varphi_w' = -\gamma C' \partial_t \varphi_w'. \tag{10}$$

The system of two curl Eqs. (7)-(8) with the two additional Eqs. (9)-(10) determines the analytical framework of the quasi-static model for the analysis of the macroscopic electromagnetic response of space-time wire media in the laboratory frame. Note that in this quasi-static model both the spatio-temporal variations of the metamaterial are homogenized and thereby the effective parameters are independent of both of space and time.

### *A. Energy conservation in space-time wire media*

Here, we derive an energy conservation theorem for space-time modulated wire media based on the quasi-static model formulation [Eqs. (7)-(10)]. The energy balance follows from applying the Poynting theorem to the coupled system described by Eqs. (7)-(8):

$$\nabla \cdot \left( \mathbf{E} \times \frac{\mathbf{B}}{\mu_0} \right) = -\partial_t \frac{1}{2}\left[ \varepsilon_0 \mathbf{E}\cdot\mathbf{E} + \frac{1}{\mu_0}\mathbf{B}\cdot\mathbf{B} \right] - \mathbf{E}\cdot\frac{1}{A_c'}\left( I'\hat{\mathbf{z}} + \gamma v C' \varphi_w' \hat{\mathbf{x}} \right). \tag{11}$$

By adding Eqs. (9)-(10) and rewriting the left-hand side in terms of a divergence, we obtain,

$$\nabla\cdot\left[\frac{1}{\gamma}I'\varphi_w'\hat{\mathbf{z}}+\left(\frac{1}{2}vL'I'^2+\frac{1}{2}vC'\varphi_w'^2\right)\hat{\mathbf{x}}\right]=-\frac{1}{2}\partial_t\left[L'I'^2+C'\varphi_w'^2\right]+I'\left(E_z+vB_y\right). \tag{12}$$

Combining Eqs. (11) and (12) results in

$$\begin{gathered}\nabla\cdot\left(\mathbf{E}\times\frac{\mathbf{B}}{\mu_0}+\frac{1}{A_c'}\frac{1}{\gamma}I'\varphi_w'\hat{\mathbf{z}}+\frac{1}{A_c'}\left(\frac{1}{2}vL'I'^2+\frac{1}{2}vC'\varphi_w'^2\right)\hat{\mathbf{x}}\right)\\=-\partial_t\frac{1}{2}\left[\varepsilon_0\mathbf{E}\cdot\mathbf{E}+\frac{1}{\mu_0}\mathbf{B}\cdot\mathbf{B}+\frac{1}{A_c'}\left(L'I'^2+C'\varphi_w'^2\right)\right]-\mathbf{E}\cdot\frac{1}{A_c'}\left(I'\hat{\mathbf{z}}+\gamma vC'\varphi_w'\hat{\mathbf{x}}\right)+\frac{1}{A_c'}I'\left(E_z+vB_y\right)\end{gathered} \tag{13}$$

which represents the energy conservation law for space-time uniaxial wire media in the form

$$\nabla\cdot\mathbf{S}+\partial_t W=-p_{loss} \tag{14}$$

with

$$\mathbf{S}=\mathbf{E}\times\frac{\mathbf{B}}{\mu_0}+\frac{1}{A_c'}\frac{1}{\gamma}I'\varphi_w'\hat{\mathbf{z}}+\frac{1}{A_c'}\left(\frac{1}{2}vL'I'^2+\frac{1}{2}vC'\varphi_w'^2\right)\hat{\mathbf{x}}, \tag{15}$$

$$W=\frac{1}{2}\left[\varepsilon_0\mathbf{E}\cdot\mathbf{E}+\frac{1}{\mu_0}\mathbf{B}\cdot\mathbf{B}+\frac{1}{A_c'}\left(L'I'^2+C'\varphi_w'^2\right)\right], \tag{16}$$

$$p_{loss}=\frac{v}{A_c'}\left(\gamma C'\varphi_w'E_x-I'B_y\right). \tag{17}$$

Here, $\mathbf{S}$ is the Poynting vector for space-time uniaxial wire media, $W$ is the density of stored energy, and $p_{loss}$ represents the net power per unit of volume absorbed by the wires in the presence of the motion (in case there is gain $p_{loss}$ is negative). The power exchange term, $p_{loss}$, reflects the non-Hermitian nature of the space-time modulation. When $v=0$, the above conservation law reduces to the result of Ref. [52]. We note for future reference that for a finite structure (such as the space-time-modulated wire medium slab considered later in the paper in Sec. III), all the field components are continuous in the co-moving frame (including the ones perpendicular to the interface) when the host dielectric is air. So, the component of the Poynting vector $\mathbf{S}$ normal to an interface (i.e., along *z*) is also continuous in the laboratory frame, because the current $I'$ must vanish at the interface [58, 60]. Thus, the energy balance is not disrupted directly by the boundary conditions at the interface.

The term $p_{loss}$ can be understood as the result of the work done by the Lorentz force acting on the charges in the "moving" metallic wires. Indeed, the effective Lorentz force is:

$$\mathbf{f}_L = \rho\mathbf{E} + \mathbf{j}\times\mathbf{B} \rightarrow f_{L,x} = \rho E_x - j_z B_y . \tag{18}$$

Therefore, using $\mathbf{j} = \frac{I'}{A'_c}\hat{\mathbf{z}} + \gamma v \rho' \hat{\mathbf{x}}$ and $\rho = \gamma\rho' = \gamma \frac{1}{A'_c} C'\varphi'_w$, we obtain:

$$f_{L,x} = \frac{1}{A'_c}\left[\gamma C'\varphi'_w E_x - I'B_y\right]. \tag{19}$$

Hence, the power absorbed by the material per unit of volume can be written as

$$p_{loss} = v f_{L,x} , \tag{20}$$

which is exactly the power transferred from the electromagnetic field to the metallic wires. In summary, the energy balance in the space-time-modulated wire medium is ruled by

$$\nabla\cdot\mathbf{S} + \partial_t W = -p_{loss} = -v f_{L,x} \tag{21}$$

with $\mathbf{S}$, $W$, and $f_{L,x}$ defined by Eqs. (15), (16), (19). We note that the above power-balance law is rather different from that in a nondispersive moving dielectric, for which it is always possible to introduce $\mathbf{S}$, $W$ such that $\nabla\cdot\mathbf{S} + \partial_t W = 0$.

The Lorentz force in time-harmonic regime is of particular interest (a time dependence in the form $e^{-i\omega t}$ is assumed, with $\omega$ real-valued),

$$\left\langle f_{L,x} \right\rangle = \frac{1}{2A'_c}\mathrm{Re}\left\{\gamma C'\varphi'_w E_x^* - I'B_y^*\right\}. \tag{22}$$

Here, $\langle ... \rangle$ denotes time-averaging in one period. Assuming $\partial_x = ik_x$, with $k_x$ real-valued, Eq. (10) yields:

$$i(\omega - vk_x)\varphi'_w = \frac{1}{\gamma C'}\partial_z I'. \tag{23}$$

Substituting Eq. (23) into Eq. (22) and taking into account that $i\omega B_y = \partial_z E_x - \partial_x E_z$ the time-averaged Lorentz force can be written as:

$$\begin{aligned} \langle f_{L,x} \rangle &= \frac{1}{2A'_c}\mathrm{Re}\left\{\frac{1}{i(\omega - vk_x)}\partial_z I' E_x^* - I' B_y^*\right\} \\ &= \frac{1}{2A'_c}\partial_z \mathrm{Re}\left\{\frac{1}{i(\omega - vk_x)} I' E_x^*\right\} - \frac{1}{2A'_c}\mathrm{Re}\left\{I'\left[\frac{-1}{(\omega - vk_x)}\left(\omega B_y + k_x E_z\right)^* + B_y^*\right]\right\} \end{aligned} \tag{24}$$

which results in

$$\langle f_{L,x} \rangle = \frac{1}{2A'_c}\partial_z \mathrm{Re}\left\{\frac{1}{i(\omega - vk_x)} I' E_x^*\right\} + \frac{1}{2A'_c}\frac{k_x}{(\omega - vk_x)}\mathrm{Re}\left\{I'\left(E_z + vB_y\right)^*\right\}. \tag{25}$$

Using the energy balance Eq. (12) for time-harmonic fields, the second term in the right-hand side can be evaluated as:

$$\begin{aligned} \mathrm{Re}\left\{I'\left(E_z + vB_y\right)^*\right\} &= \nabla \cdot \mathrm{Re}\left[\frac{1}{\gamma} I'^* \varphi'_w \hat{\mathbf{z}} + \left(\frac{1}{2}vL'|I'|^2 + \frac{1}{2}vC'|\varphi'_w|^2\right)\hat{\mathbf{x}}\right] \\ &= \frac{\partial}{\partial z}\mathrm{Re}\left\{\frac{1}{\gamma} I'^* \varphi'_w\right\}. \end{aligned} \tag{26}$$

In Eq. (26) we took into account that $|I'|^2, |\varphi'_w|^2$ are independent of *x*. Thus, the Lorentz force given by Eq. (25) can finally be expressed as:

$$\langle f_{L,x} \rangle = \frac{1}{2A'_c}\frac{\partial}{\partial z}\mathrm{Re}\left\{\frac{1}{i(\omega - vk_x)} I' E_x^*\right\} + \frac{1}{2A'_c}\frac{k_x}{(\omega - vk_x)}\frac{\partial}{\partial z}\mathrm{Re}\left\{\frac{1}{\gamma} I'^* \varphi'_w\right\}. \tag{27}$$

Hence, for a space-time wire medium slab, as will be considered in Sec. III in the context of wave scattering, the total Lorentz force [Eq. (27)] vanishes after being integrated over the wire medium slab because the current satisfies the (additional) boundary condition $I'=0$ at the interfaces [58,60].

In summary, in time-harmonic regime (with $\omega$ real-valued) and for fields such that $\partial_x = ik_x$ the net power transferred to the wire medium slab vanishes, $\int dz \langle p_{loss} \rangle = 0$. Combining this result with Eq. (21), it follows that the flux of the Poynting vector at the two interfaces ( $z = const.$, top and bottom) with air is necessarily identical. Thus, despite the non-Hermitian character of the space-time modulation and the nontrivial local exchange of power between the field and the space-time wire medium, the net energy flux is conserved for this scattering configuration.

### *B. Effective response of space-time wire media*

Next, we derive the Landau-Lifshitz effective permittivity tensor for the space-time-modulated wire media based on the quasi-static model in the beginning of this section. In the Fourier transform domain, Eqs. (9) and (10) for the current and the additional potential in the laboratory frame lead to

$$\frac{1}{\gamma} ik_z \varphi'_w - iL'\left(\omega - k_x v\right) I' = E_z + vB_y, \tag{28}$$

$$\frac{1}{\gamma} ik_z I' - iC'\left(\omega - k_x v\right) \varphi'_w = 0. \tag{29}$$

Hence, $\varphi'_w = \frac{1}{\gamma} \frac{k_z}{C'\left(\omega - k_x v\right)} I'$, so that

$$I' = -i\left[\gamma\left(\omega - k_x v\right)\right] \frac{C'}{k_z^2 - L'C'\left[\gamma\left(\omega - k_x v\right)\right]^2} \gamma\left(E_z + vB_y\right), \tag{30}$$

$$\varphi_w' = -\frac{ik_z}{k_z^2 - L'C'\left[\gamma\left(\omega - k_x v\right)\right]^2}\gamma\left(E_z + vB_y\right). \tag{31}$$

Substituting Eqs. (30), (31) into Eq. (8), the equivalent polarization vector $\mathbf{P}$ can be obtained as follows:

$$\begin{aligned}\mathbf{P} &= \frac{1}{-i\omega}\frac{1}{A_c'}\left(I'\hat{\mathbf{z}} + \gamma v C'\varphi_w'\hat{\mathbf{x}}\right) \\ &= -\frac{1}{A_c'}\frac{C'}{L'C'\left[\gamma\left(\omega - k_x v\right)\right]^2 - k_z^2}\left[\frac{\left(\omega - k_x v\right)}{\omega}\hat{\mathbf{z}} + \frac{k_z}{\omega}v\hat{\mathbf{x}}\right]\gamma^2\left(E_z + vB_y\right)\end{aligned}. \tag{32}$$

Since, $\mathbf{k}\times\mathbf{E} = \omega\mathbf{B}$, leading to $\omega B_y = k_z E_x - k_x E_z$, Eq. (32) becomes

$$\mathbf{P} = -\frac{1}{A_c'}\frac{C'}{L'C'\left[\gamma\left(\omega - k_x v\right)\right]^2 - k_z^2}\left[\left(1 - \frac{k_x}{\omega}v\right)\hat{\mathbf{z}} + v\frac{k_z}{\omega}\hat{\mathbf{x}}\right]\gamma^2\left(\left(1 - \frac{k_x}{\omega}v\right)E_z + v\frac{k_z}{\omega}E_x\right). \tag{33}$$

Equations (8) and (33) allow us to define the Landau-Lifshitz effective permittivity tensor for space-time uniaxial wire media (Fig. 1(c)):

$$\frac{\bar{\varepsilon}}{\varepsilon_0} = \frac{1}{\varepsilon_0}\begin{pmatrix}\varepsilon_{xx} & 0 & \varepsilon_{xz} \\ 0 & \varepsilon_{yy} & 0 \\ \varepsilon_{zx} & 0 & \varepsilon_{zz}\end{pmatrix} \tag{34}$$

with

$$\frac{\varepsilon_{xx}}{\varepsilon_0} = 1 - \frac{1}{A_c'}\frac{C'/\varepsilon_0}{L'C'\left[\gamma\left(\omega - k_x v\right)\right]^2 - k_z^2}\left(v\frac{k_z}{\omega}\right)^2\gamma^2, \tag{35a}$$

$$\frac{\varepsilon_{yy}}{\varepsilon_0} = 1, \tag{35b}$$

$$\frac{\varepsilon_{zz}}{\varepsilon_0} = 1 - \frac{1}{A_c'}\frac{C'/\varepsilon_0}{L'C'\left[\gamma\left(\omega - k_x v\right)\right]^2 - k_z^2}\left(1 - \frac{k_x}{\omega}v\right)^2\gamma^2, \tag{35c}$$

$$\frac{\varepsilon_{xz}}{\varepsilon_0} = \frac{\varepsilon_{zx}}{\varepsilon_0} = -\frac{1}{A_c'}\frac{C'/\varepsilon_0}{L'C'\left[\gamma\left(\omega - k_x v\right)\right]^2 - k_z^2}\left(1 - \frac{k_x}{\omega}v\right)\left(v\frac{k_z}{\omega}\right)\gamma^2. \tag{35d}$$

The tensor has a pole when $L'C'\left[\gamma\left(\omega-k_x v\right)\right]^2-k_z^2=0$, which is inherited from the strong spatial dispersion of the static wire-medium.

For frequency and wave vector real-valued, the Landau-Lifshitz permittivity tensor [Eqs. 34, (35a)-(35d)] is real-valued and symmetric. However, the effective response violates the reciprocity condition $\bar{\varepsilon}\left(\omega,\mathbf{k}\right)=\bar{\varepsilon}^{\mathrm{T}}\left(\omega,-\mathbf{k}\right)$ [64], demonstrating that the medium is nonreciprocal.

In the Landau–Lifshitz representation, the complete material response is incorporated into an equivalent nonlocal permittivity tensor. In particular, the terms that are linear in the components of the wave vector account for the effective bianisotropic, or magnetoelectric-coupling, response. In Appendix A, we obtain the effective response in the laboratory frame directly from the effective response in the co-moving frame using an inverse Lorentz transformation [13]. In this alternative formulation, the effective medium response is characterized by permittivity, permeability, and magnetoelectric coupling tensors [13]. This alternative description makes clear that the synthetic motion makes the laboratory-frame response of the spatio-temporal modulated wire medium bianisotropic. Because the wire medium is anisotropic in the co-moving frame, its magnetoelectric tensors do not exhibit the antisymmetric form characteristic of an isotropic moving medium. Instead, they exhibit a different tensorial structure consistent with a uniaxial permittivity response.

The Landau-Lifshitz permittivity tensor satisfies (for $\left(\omega,\mathbf{k}\right)$ real-valued)

$$\bar{\varepsilon}\left(\omega,\mathbf{k}\right)=\bar{\varepsilon}^{\dagger}\left(\omega,\mathbf{k}\right), \tag{36}$$

which is a necessary (but not sufficient) condition so that the spatially dispersive response of the bulk material is conservative [64]. Note that Eq. (36) implies that the power dissipated vanishes for individual bulk plane waves with real-valued frequency. However, this condition does not, in general, guarantee a stable electromagnetic response. Indeed, moving systems may satisfy an analogous relation while still developing electromagnetic instabilities in configurations involving two or more bodies that do not share a common co-moving frame for $\omega$ complex [65]. In this article, we restrict our attention to a single spatiotemporally modulated slab. This system is always stable. The reason is straightforward: any instability in the laboratory frame would imply a corresponding instability in the co-moving frame. Such an instability is not physically possible, since in the co-moving frame the material response is manifestly stable and corresponds to that of a regular wire medium.

### C. *Isofrequency contours and Fresnel drag*

The isofrequency contours of the homogenized space-time wire medium can be found in a standard way by solving the secular dispersion equation $\det\left(k^2\mathbf{1}-\mathbf{k}\otimes\mathbf{k}-\left(\frac{\omega}{c}\right)^2\bar{\varepsilon}(\omega,\mathbf{k})\right)=0$. In the absence of a space-time modulation, it is well known that the homogenized wire medium (with the wires oriented along the $z-$direction as in Fig. 1(a)) supports three different plane-wave modes: two extraordinary modes due to spatial dispersion, specifically a transverse electromagnetic (TEM) mode and a transverse magnetic (TM-z) mode, and an ordinary transverse electric (TE-z) mode [50]. The isofrequency surfaces of these modes are determined by the following dispersion equations in the static (co-moving) frame:

$$\left(\frac{\omega'}{c}\right)^2 = k_x'^2 + k_y'^2 + k_z'^2, \qquad \text{(TE mode)} \tag{37}$$

$$\left(\frac{\omega'}{c}\right)^2 = k_p^2 + k_x'^2 + k_y'^2 + k_z'^2, \qquad \text{(TM mode)} \tag{38}$$

$$\frac{\omega'}{c} = \pm k_z'. \qquad \text{(TEM mode)} \tag{39}$$

For a non-trivial modulation velocity $v$, these dispersions become:

$$\left(\frac{\omega}{c}\right)^2 = k_x^2 + k_y^2 + k_z^2, \qquad \text{(TE mode)} \qquad (40)$$

$$\left(\frac{\omega}{c}\right)^2 = k_p^2 + k_x^2 + k_y^2 + k_z^2. \qquad \text{(TM mode)} \qquad (41)$$

$$\omega = v k_x \pm \frac{c}{\gamma} k_z, \qquad \text{(TEM mode).} \qquad (42)$$

Rather remarkably, the dispersions of the TE and TM modes are velocity independent. A simple way to explain this result is to note that the co-moving and laboratory frame dispersions can be linked directly through a relativistic Doppler transformation [62],

$$\omega' = \gamma\left(\omega - v k_x\right), \;\; k_x' = \gamma\left(k_x - \omega v / c^2\right), \;\; k_y' = k_y, \;\; k_z' = k_z. \qquad (43)$$

As $\left(\frac{\omega}{c}, \mathbf{k}\right)$ transforms as a 4-vector and the Minkowski product of two 4-vectors is a Lorentz scalar, it is evident that $\left(\frac{\omega}{c}\right)^2 - \mathbf{k}\cdot\mathbf{k}$ must remain unchanged under a Lorentz transformation, in agreement with the above results.

Clearly, from Eqs. (40)-(41), the isofrequency surfaces of the TE and TM modes are independent of the modulation velocity $v$, and thereby are exactly spherical surfaces. In contrast, the dispersion of the TEM mode [Eq. (42)] corresponds to two tilted intersecting planes in the $k$-space, as it is illustrated in Fig. 2. The isofrequency contours are plotted in the $(k_x, k_z)$-plane for the TEM and TE modes at the frequencies of 1 GHz (Fig. 2(a)) and 2 GHz (Fig. 2(b)), respectively, (with the dispersion given by Eqs. (40) and (42)). It is implicit that $k_y = 0$. For simplicity, we omit the TM mode because it is an evanescent mode below the plasma frequency of the wire medium.

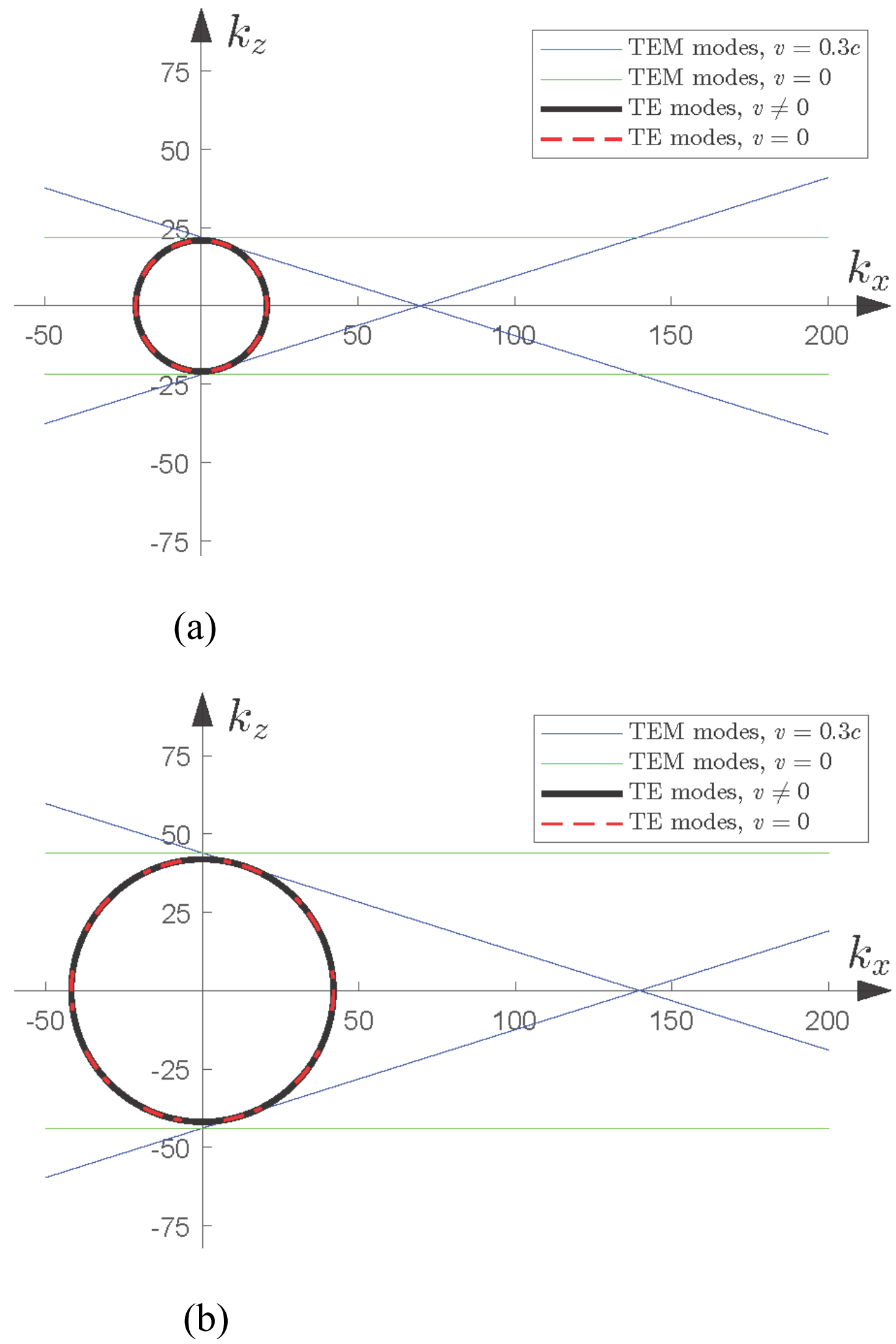

Fig. 2. Isofrequency contours for the TEM and TE modes of the static and moving wire media at the frequencies of (a) 1 GHz and (b) 2 GHz. Note that these isofrequency contours are independent of the structural parameters of the wire medium. The units of the wave numbers $k_x$ and $k_z$ are rad/m.

The isofrequency contours in Fig. 2 are plotted for both the forward ($k_z > 0$) and backward ($k_z < 0$) modes. As expected, in the static regime ($v = 0$) the TEM modes (forward and backward) propagate along the wires in the $z-$direction independent of the wavenumber $k_x$ (two parallel contours in Fig. 2). In the presence of the motion ($v > 0$), this symmetry is broken resulting in two tilted intersecting isofrequency contours. Interestingly, at the intersection point of forward and backward solutions of the TEM modes (where $k_z = 0$ at $k_x = \omega / v$) the solutions interchange from forward to backward and vice versa. The intersection point determines a Cherenkov-like threshold for waves with a certain $k_x$, because the phase velocity along $x$, $\omega / k_x$, becomes smaller than $v$ after the threshold. The threshold is also related to a transition from passive to a gain response [65]. For example, gain-like effects usually come from the interaction of the parts of the branch with $k_x > \omega / v$, which correspond to evanescent waves when the system is coupled to free space. As already noted, unlocking gain in this system requires the interaction of systems with different modulation velocities [65]. It is interesting to note that isofrequency contour of the TEM wave remains tangent (it has the same local slope) to the isofrequency contour of the TE wave, even for large modulation velocities.

The key observation is that the dispersion of the TEM modes in the space-time modulated wire medium depends on the wavenumber $k_x$, different from the static case. Hence the TEM modes no longer propagate along the wires in the $z-$direction, but at some angle in the $(x,z)$-plane depending on the modulation velocity $v$. The group velocity is determined as follows,

$$\mathbf{v}_g = \nabla_{\mathbf{k}} \omega = v\hat{\mathbf{x}} \pm \frac{c}{\gamma} \hat{\mathbf{z}}. \tag{44}$$

It should be noted that $\left|\mathbf{v}_g\right| = \sqrt{v^2 + \frac{c^2}{\gamma^2}} = c$. The component along the $x$-direction can be attributed to a Fresnel-Fizeau-drag like effect: the TEM wave is dragged by the synthetic motion of the metamaterial (its velocity along the $z$-direction is also reduced to ensure $\left|\mathbf{v}_g\right| = c$). This contrasts with the TE and TM waves whose dispersion is unaffected by the synthetic motion. The wire medium therefore provides a unique platform for investigating Fresnel-drag phenomena. In particular, its response differs substantially from the conventional picture, in which the isofrequency contour is represented by an ellipse displaced in momentum space.

The Fresnel drag controls the lateral shifts (Goos-Hänchen-like shifts) discussed in Section III.A in the context of wave scattering by a finite-thickness space-time-modulated wire medium slab. Specifically, as the beam transverses a distance *d* along *z* it should suffer the lateral shift:

$$\Delta = v_{g,x}\delta t \text{ with } \delta t \approx \frac{d}{v_{g,z}}. \quad (45)$$

In the above, $v_{g,x}, v_{g,z}$ are the *x* and *z* components of the group velocity of the relevant mode. For the space-time modulated wire medium, we obtain with the help of Eq. (44),

$$\Delta = v\frac{\gamma d}{c}. \quad (46)$$

Note that the shift is independent of the incident angle because $\mathbf{v}_g$ is independent of the wave vector. In the reflection case, the shift needs to be multiplied by a factor of 2 to account for the round-trip effect. Evidently, this calculation neglects multiple reflections, which may affect the result.

It is important to note that the discussed Fresnel drag is not specific to the wire medium and can be realized with arrays of thin metallic scatterers of arbitrary shape, such as metallic disks [66], contained in the *xz*-plane. In fact, the response of such systems remains Lorentz invariant. Hence, their effective response in the co-moving frame can be described simply by permittivity and permeability tensors [66], which, for subwavelength scatterers, may be assumed to be approximately local and weakly dependent on frequency. Similar to the wire medium, the corresponding moving-medium-like response can be realized by modulating only the metallic response, without requiring an explicit modulation of the permeability. Since the group velocity in the laboratory frame is related to that in the co-moving frame through the relativistic addition of velocities [12], it follows that such systems will generally exhibit analogous Fresnel-like phenomena. The key difference with respect to wire media is that, in these systems, the co-moving-frame isofrequency contours tend to be similar to those of the background material and are therefore less affected by the synthetic motion. In contrast, the isofrequency contour of the TEM mode of the wire medium already differs radically from that of the homogeneous background, potentially leading to a stronger Fresnel-like effect.

## III. Scattering of Plane Waves from Space-Time Wire Medium Slab

In this Section we study scattering of electromagnetic plane waves from a space-time-modulated wire medium slab. The geometry of the wire medium slab in the laboratory frame, with a modulation velocity of the form $\mathbf{v} = \hat{\mathbf{x}} v$, is shown in Fig. 3. The incident plane wave propagates in air and in general, represents a superposition of transverse magnetic (TM) and transverse electric (TE) waves with the transverse wave numbers $k_x$ and $k_y$. The scattering problem is formulated in terms of incident, reflected, and transmitted electric fields, related through reflection $\mathbf{R}$ and transmission $\mathbf{T}$ matrices. The $\mathbf{R}$ ($\mathbf{T}$) matrix relates the transverse components ($x$ and $y$) of the incident field with the transverse components of the reflected (transmitted) field. The direction of incidence is arbitrary.

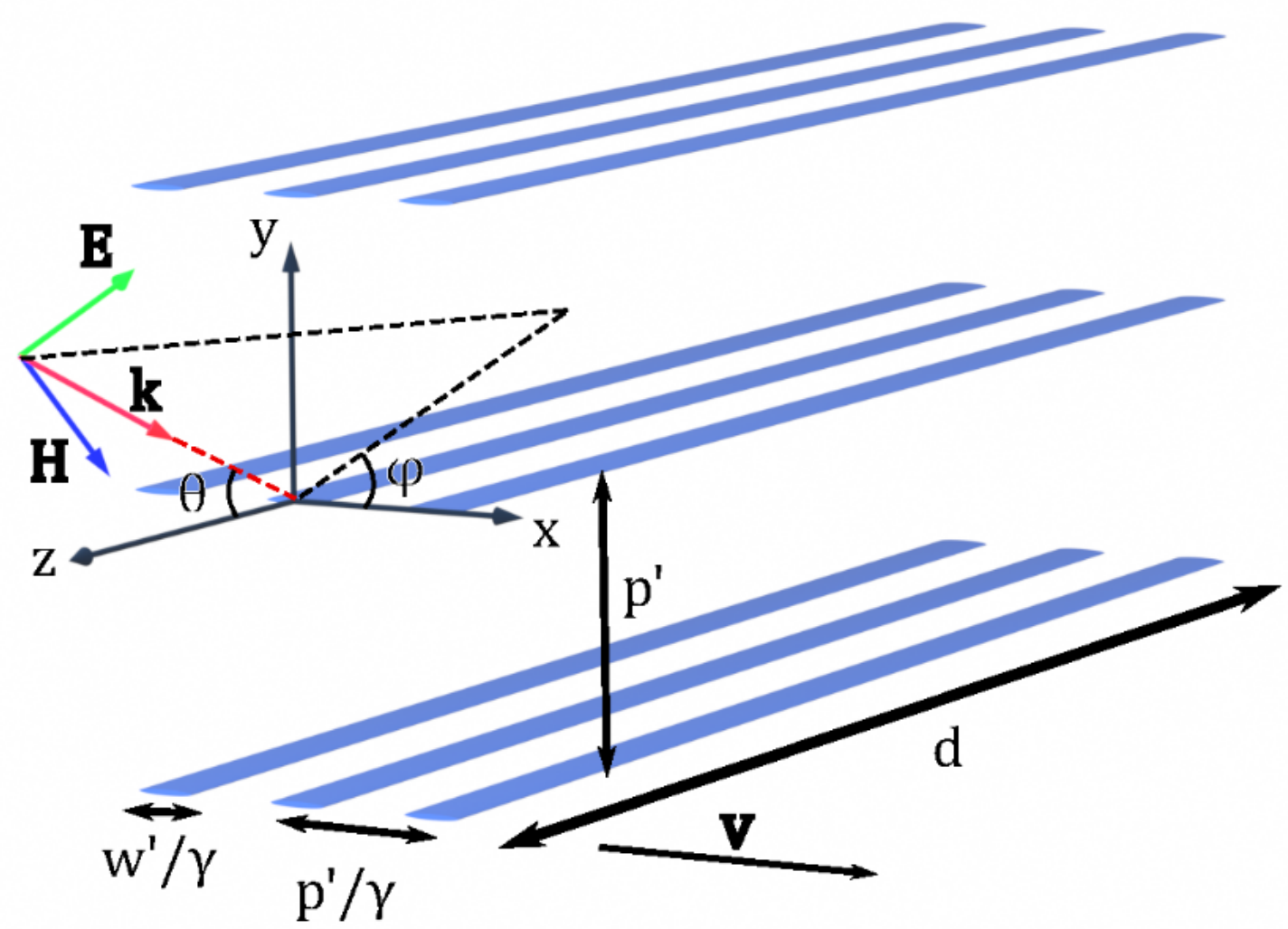

**Fig. 3** Geometry of a moving wire medium slab of thickness $d$ in the laboratory frame with the arbitrarily incident plane wave. The wires are approximated by thin strips, with $w = w'/\gamma$ and $p = p'/\gamma$ in the direction of the motion.

Although the laboratory frame reflection and transmission matrices $\mathbf{R}$ and $\mathbf{T}$ could be obtained directly from the homogenization formalism of Sec. II, it is more convenient to determine them by combining the analytical solution previously established for the static wire-medium slab with the appropriate Lorentz transformations between the co-moving and laboratory frames. Specifically, $\mathbf{R}$ and $\mathbf{T}$, respectively, are related to the co-moving frame reflection and transmission matrices, $\mathbf{R}_{\text{co}}$ and $\mathbf{T}_{\text{co}}$, as follows [65]:

$$\mathbf{R}\left(\omega,k_x,k_y\right)=\left(\mathbf{I}_{2\times2}+\mathbf{A}\right)^{-1}\cdot\mathbf{R}_{\text{co}}\left(\omega',k'_x,k'_y\right)\cdot\left(\mathbf{I}_{2\times2}+\mathbf{A}\right), \tag{47a}$$

$$\mathbf{T}\left(\omega,k_x,k_y\right)=\left(\mathbf{I}_{2\times2}+\mathbf{A}\right)^{-1}\cdot\mathbf{T}_{\text{co}}\left(\omega',k'_x,k'_y\right)\cdot\left(\mathbf{I}_{2\times2}+\mathbf{A}\right), \tag{47b}$$

where

$$\mathbf{A}=\gamma\begin{pmatrix}0 & 0\\ v\dfrac{k_y}{\omega} & 1-\dfrac{1}{\gamma}-v\dfrac{k_x}{\omega}\end{pmatrix}. \tag{47c}$$

It can be shown [65] that the co-moving frame reflection and transmission coefficients for TM and TE polarized plane waves, $R_{\text{TM,co}}$, $R_{\text{TE,co}}$ and $T_{\text{TM,co}}$, $T_{\text{TE,co}}$, represent the eigenvalues of $\mathbf{R}_{\text{co}}$ and $\mathbf{T}_{\text{co}}$. Moreover, $\mathbf{k}'_t / |k'_t|$ is the eigenvector for TM polarization (with the electric field of the incident wave parallel to the plane of incidence), and $\mathbf{k}'_{tn} / |k'_{tn}|$ is the eigenvector for TE polarization (with the electric field of the incident wave perpendicular to the plane of incidence). Then, based on the spectral decomposition [67] the reflection and transmission matrices in the co-moving frame can be expressed by:

$$\mathbf{R}_{\text{co}} = \frac{1}{k_t'^2}\left[R_{\text{TM,co}}\left(\omega', k'_x, k'_y\right)\mathbf{k}'_t \otimes \mathbf{k}'_t + R_{\text{TE,co}}\left(\omega', k'_x, k'_y\right)\mathbf{k}'_{tn} \otimes \mathbf{k}'_{tn}\right],$$

(48a)

$$\mathbf{T}_{\text{co}} = \frac{1}{k_t'^2}\left[T_{\text{TM,co}}\left(\omega', k'_x, k'_y\right)\mathbf{k}'_t \otimes \mathbf{k}'_t + T_{\text{TE,co}}\left(\omega', k'_x, k'_y\right)\mathbf{k}'_{tn} \otimes \mathbf{k}'_{tn}\right], \tag{48b}$$

where $\mathbf{k}'_t = \left(k'_x, k'_y\right)$ and $\mathbf{k}'_{tn} = \left(-k'_y, k'_x\right)$. The parameters $(\omega', k'_x, k'_y)$ are related to those in the laboratory frame by a relativistic Doppler transformation [Eq. (43)]. The expressions for the reflection and transmission coefficients, $R_{\text{TM,co}}$ and $T_{\text{TM,co}}$, for a wire medium slab in the static (co-moving) frame for the TM-polarized obliquely incident plane wave have been obtained in [32] and are documented in Appendix B.

The co-polarized and cross-polarized reflection and transmission coefficients for TM and TE incident plane waves are determined by:

$$r_{\text{TM,TM}} = \frac{1}{k_t^2}\mathbf{k}_t^* \cdot \mathbf{R} \cdot \mathbf{k}_t = \frac{1}{k_x^2 + k_y^2}\begin{pmatrix} k_x & k_y \end{pmatrix} \cdot \mathbf{R} \cdot \begin{pmatrix} k_x \\ k_y \end{pmatrix}, \tag{49a}$$

$$r_{\text{TM,TE}} = \frac{1}{k_t^2}\mathbf{k}_t^* \cdot \mathbf{R} \cdot \mathbf{k}_{tn} = \frac{1}{k_x^2 + k_y^2}\begin{pmatrix} k_x & k_y \end{pmatrix} \cdot \mathbf{R} \cdot \begin{pmatrix} -k_y \\ k_x \end{pmatrix}, \tag{49b}$$

$$r_{\mathrm{TE,TM}} = \frac{1}{k_t^2}\mathbf{k}_{tn}^* \cdot \mathbf{R} \cdot \mathbf{k}_t = \frac{1}{k_x^2 + k_y^2}\begin{pmatrix} -k_y & k_x \end{pmatrix} \cdot \mathbf{R} \cdot \begin{pmatrix} k_x \\ k_y \end{pmatrix}, \tag{49c}$$

$$r_{\mathrm{TE,TE}} = \frac{1}{k_t^2}\mathbf{k}_{tn}^* \cdot \mathbf{R} \cdot \mathbf{k}_{tn} = \frac{1}{k_x^2 + k_y^2}\begin{pmatrix} -k_y & k_x \end{pmatrix} \cdot \mathbf{R} \cdot \begin{pmatrix} -k_y \\ k_x \end{pmatrix}, \tag{49d}$$

with similar expressions for the transmission coefficients $t_{\mathrm{TM,TM}}, t_{\mathrm{TM,TE}}, t_{\mathrm{TE,TM}}, t_{\mathrm{TE,TE}}$. A global energy-balance law [Eq. (C4)] is derived in Appendix C for propagating plane waves, and has been numerically validated (not shown). This result confirms that for real-valued frequencies the flux of energy through the wire medium interfaces is conserved, in agreement with Sect. II.A.

Figures 4 and 5 represent the magnitude of the reflection and transmission coefficients for TM and TE obliquely incident plane waves on the space-time-modulated wire medium slab with the following parameters: slab thickness $d = 5$ mm, strip width $w = 0.1$ mm, period of strips $p = 1$ mm, and the angle of incidence $\varphi = 60^\circ$. The results in Figs. 4 and 5 clearly demonstrate that the synthetic motion of the wire medium leads to a cross-coupling between TE and TM polarizations. This is attributed to the bianisotropic response of space-time-modulated wire medium slab. We should also point out that even though $R_{\mathrm{TE,co}}(\omega', k_x', k_y') = 0$ in the co-moving frame, the reflection coefficients $r_{\mathrm{TE,TE}}$, $r_{\mathrm{TE,TM}}$, and $r_{\mathrm{TM,TE}}$ are non-zero in the laboratory frame, and strongly depend on the modulation velocity $v$.

The synthetic motion also leads to non-reciprocity, which is manifested in an asymmetric coupling between different incident directions. Specifically, reciprocity requires that $r_{\mathrm{TE,TE}}(k_x, k_y) = r_{\mathrm{TE,TE}}(-k_x, -k_y)$, $r_{\mathrm{TM,TM}}(k_x, k_y) = r_{\mathrm{TM,TM}}(-k_x, -k_y)$, and $r_{\mathrm{TE,TM}}(k_x, k_y) = r_{\mathrm{TM,TE}}(-k_x, -k_y)$. Note that the incident transverse wave vector $(k_x, k_y)$ is mapped onto some incidence angle $(\theta,\varphi)$, whereas the incident wave vector $(-k_x,-k_y)$ is mapped onto $(\theta,\varphi+180^{\circ})$. The two directions are linked by Snell's reflection law. For simplicity, we shall refer to $(\theta,\varphi)$ as simply $\theta$ and to $(\theta,\varphi+180^{\circ})$ as $-\theta$. Reciprocity also imposes a constraint on the transmission matrix involving illumination from opposite sides of the slab, which is omitted here for conciseness.

Next, we demonstrate numerically the reciprocity violation, by studying the scattering by the space-time-modulated wire medium slab for the angles of incidence $\pm\theta$. In Fig. 4 it can be seen that for $v=0$ (black line) the reflection coefficient $r_{\mathrm{TM,TM}}$ satisfies the reciprocity constraints (note that the remaining reflection coefficients are identically zero when $v=0$). In contrast, for a finite modulation speed there is strong sensitivity to the incidence angle $\pm\theta$ (red and green lines, respectively), confirming the nonreciprocal response of the material. Figure 5 shows a corresponding asymmetry in the transmission coefficients.

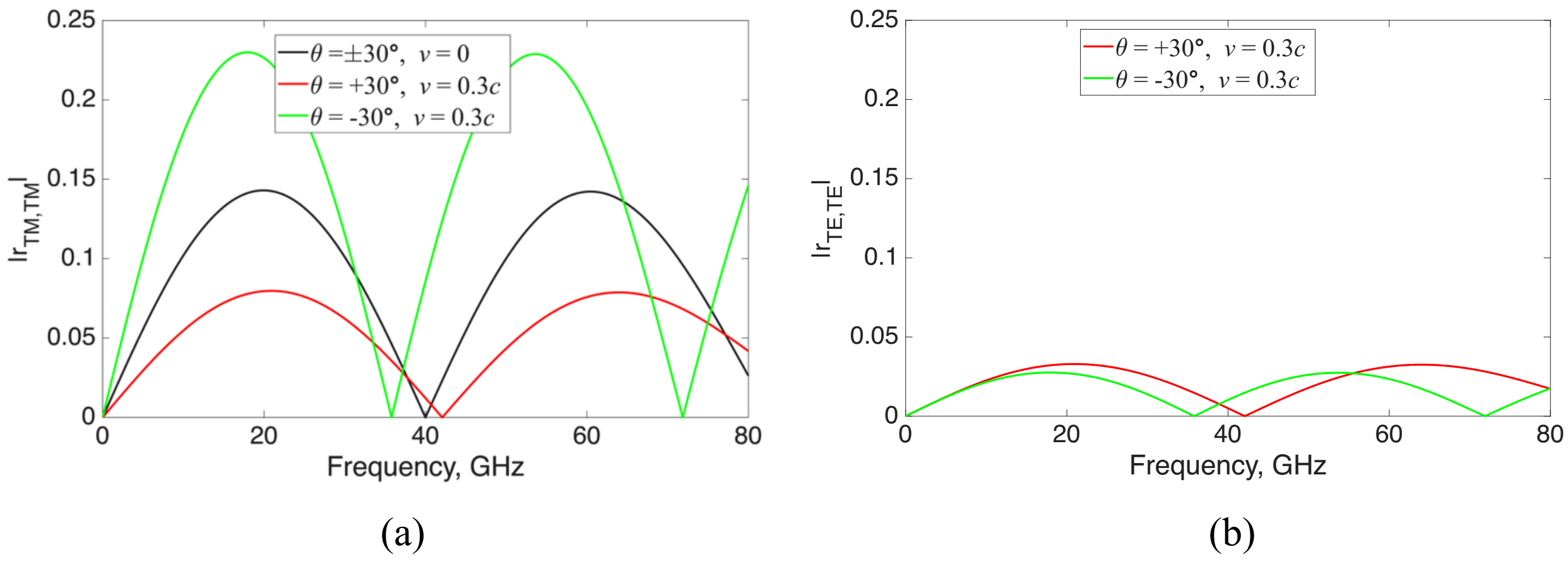

(a) (b)

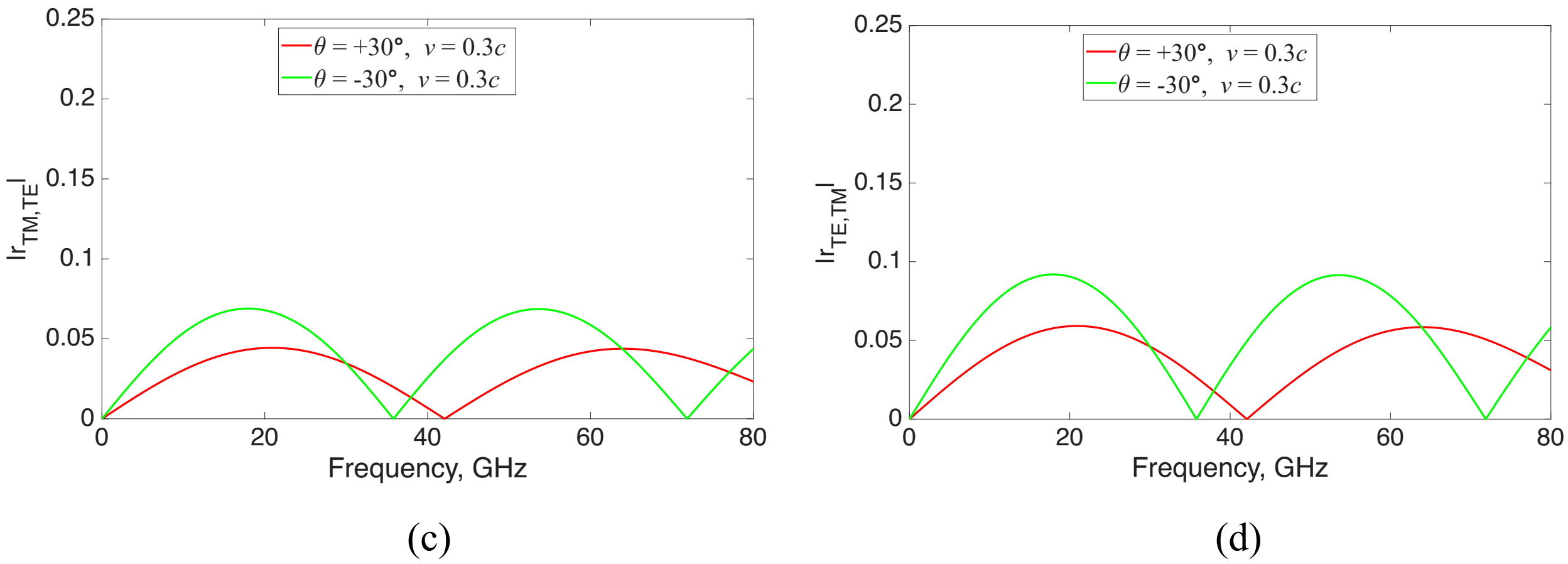

**Fig. 4** Magnitude of the reflection coefficients in the laboratory frame for space-time-modulated wire medium slab with the obliquely incident TM and TE polarized plane waves at $\pm\theta$.

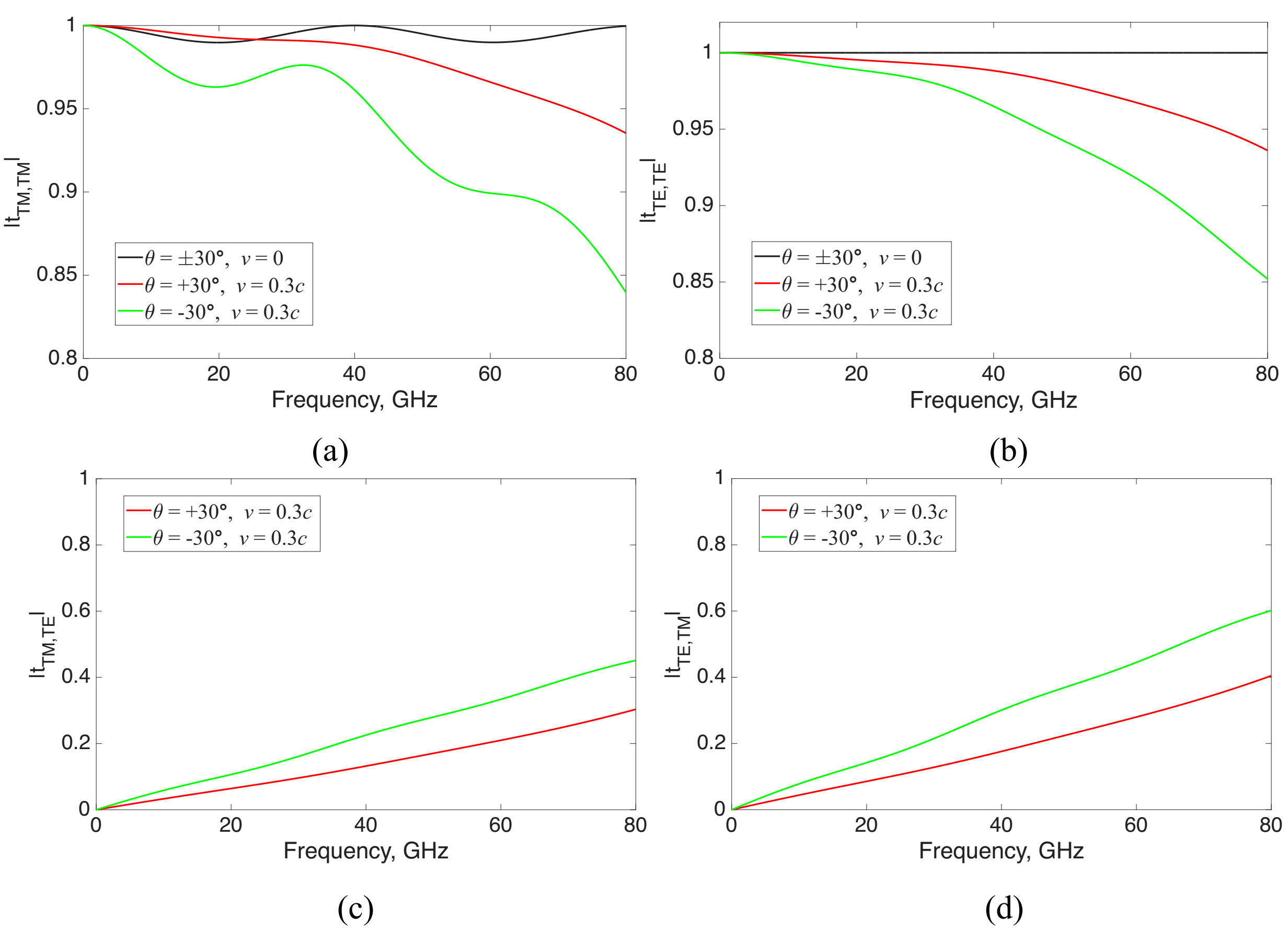

**Fig. 5** Magnitude of the transmission coefficients in the laboratory frame for space-time-modulated wire medium slab with the obliquely incident TM and TE polarized plane waves at $\pm\theta$.

### *A. Spatial reflection and transmission shifts for a moving wire medium slab*

Using the obtained analytical expressions for the reflection and transmission matrices $\mathbf{R}$ and $\mathbf{T}$, in what follows we study the Goos-Hänchen-like shift suffered by a beam incident on the space-time modulated wire medium slab. The Goos-Hänchen shift is the displacement of the reflected/transmitted beam with respect to the perfectly specular reflection/transmission. For the case of the transmission shift (TS), we consider a wire medium slab standing in air, similar to the previous examples. In contrast, for the case of the reflection shift (RS), we assume that the wire medium slab is backed by a ground plane.

The lateral shifts are studied for TM-polarized waves based on either the reflection coefficient $r_{\mathrm{TM,TM}}$ [Eq. (49a)] or transmission coefficient $t_{\mathrm{TM,TM}}$ (with a similar expression as Eq. (49a) in terms of the transmission matrix $\mathbf{T}$), depending on the type of shift considered. The reflection and transmission shifts are defined with respect to the direction of the motion ($x-$axis) for $\varphi = 0^{\circ}$. The lateral shifts for the reflected field (RS) and for the transmitted field (TS) can be evaluated from the phase of the reflection and transmission coefficients as follows [17], [68]:

$$\mathrm{RS} = -\frac{\partial \arg(r_{\mathrm{TM,TM}})}{\partial k_x} = -\frac{1}{k_0}\frac{\partial \arg(r_{\mathrm{TM,TM}})}{\partial \sin(\theta)}, \tag{50}$$

$$\mathrm{TS} = -\frac{\partial \arg(t_{\mathrm{TM,TM}})}{\partial k_x} = -\frac{1}{k_0}\frac{\partial \arg(t_{\mathrm{TM,TM}})}{\partial \sin(\theta)}. \tag{51}$$

where $k_0 = \omega / c$. The numerical results for RS and TS in Fig. 6 have been obtained for the following parameters of space-time wire medium slab: strip width $w = 4\ \mathrm{mm}$, period of strips $p = 1$ cm, and frequency $f = 1$ GHz. In these simulations, we consider different values for the thickness $d$ of wire medium slab and for the modulation velocity $v$.

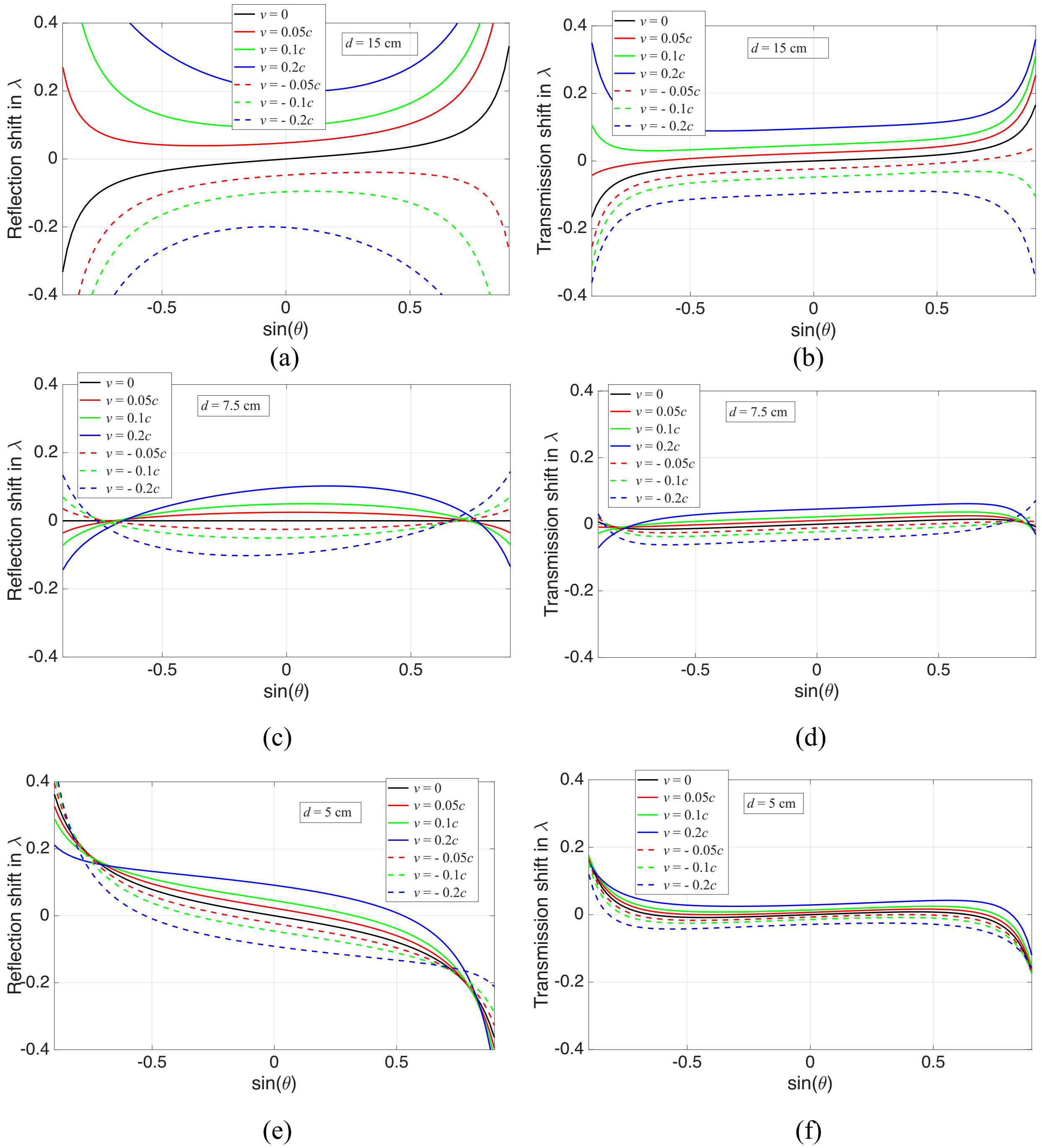

**Fig. 6** Goos-Hänchen reflection shift (RS) and the transmission shift (TS) for a space-time modulated wire medium slab parameterized by the velocity $v$; (a)-(b) slab thickness $d = 15$ cm ($\lambda/2$ at 1 GHz for $v = 0$), (c)-(d) slab thickness $d = 7.5$ cm ($\lambda/4$ at 1 GHz for $v = 0$), and (e)-(f) slab thickness $d = 5$ cm ($\lambda/6$ at 1 GHz for $v = 0$).

It can be seen in Fig. 6 that the reflection and transmission shifts for the static (co-moving, $v=0$) wire medium slab are the odd functions of $\theta$, such that the RS ($\Delta_{\rm RS}$) and TS ($\Delta_{\rm TS}$) satisfy $\Delta_{\rm RS}(\theta)=-\Delta_{\rm RS}(-\theta)$ and $\Delta_{\rm TS}(\theta)=-\Delta_{\rm TS}(-\theta)$. On the other hand, for the space-time-modulated wire medium slab with $v\neq 0$, RS and TS are the odd functions of both $\theta$ and $v$, $\Delta_{\rm RS}(\theta,v)=-\Delta_{\rm RS}(-\theta,-v)$, $\Delta_{\rm TS}(\theta,v)=-\Delta_{\rm TS}(-\theta,-v)$, where the negative velocity corresponds to the motion in the $-x-$direction.

For long wires (see the case of $d=15$ cm ($\lambda/2$ at 1 GHz) in Fig. 6 (a),(b), the RS and TS are positive for a positive $v$ (and vice versa for negative $v$). The shifts are well approximated by the formulas $\Delta_{\rm RS}\approx(2d/c)\gamma v$ and $\Delta_{\rm TS}\approx(d/c)\gamma v$ [Eq. (46)], confirming that they arise from the synthetic Fresnel-Fizeau drag. This is due to propagation of the extraordinary TEM wave in the wire medium primarily along the wires, which is dragged in the direction of the synthetic motion. For shorter wires in Fig. 6 (c),(d) multiple reflections and the effect of the extraordinary evanescent TM mode influence the lateral shifts, and the discussed approximations are no longer accurate. Figure 6(c) shows that for the case of a quarter-wavelength grounded wire medium slab in the co-moving frame ($v=0$), the RS is zero for all the incidence angles. This is due to a perfect magnetic conductor condition with the zero reflection phase for all incidence angles (note that the reflection coefficient $R_{\rm TM,co}(\omega',k'_x,k'_y)=1$ in Eq. (B3)). However, for a "moving" grounded wire medium slab the Doppler transformation mixes frequency with the wave vector. Consequently, the RS depends on the modulation velocity and on the incidence angle in the laboratory frame. For a wire medium slab with very short wires (Fig. 6 (e),(f)) the Fresnel drag effect is very much weakened, eventually leading to null shifts as $d\to 0$.

In summary, the results in Fig. 6 demonstrate how the space-time-modulated wire medium slab effectively drags the incident beam, producing a significant Fresnel-Fizeau drag-like effect with lateral shifts roughly proportional to the synthetic velocity.

## IV. Conclusion

Summarizing the study in this paper, we have demonstrated that space–time modulation of metallic-like responses can produce synthetic Fresnel–Fizeau drag in the homogenization limit without requiring the simultaneous modulation of permittivity and permeability. In this way, the proposed approach overcomes a key limitation of previous homogenization theories of moving-medium analogues. The wire medium provides a particularly favorable realization of this principle, as its intrinsic anisotropy can enhance the synthetic-drag effect.

Using the equivalence between the considered space-time modulated system and a moving crystal, we have developed a quasi-static analytical framework for the interaction of electromagnetic waves and space-time-modulated wire media. In particular, we have derived the Landau-Lifshitz effective permittivity tensor for a homogenized uniaxial space-time wire medium and an energy-balance law. Our analysis shows that while energy can be exchanged locally between the wave and the modulation circuit, in typical scattering problems the net energy exchange vanishes, and global energy conservation is therefore preserved. We have also obtained the constitutive parameters of the "moving" wire medium in terms of an effective material matrix comprising permittivity, permeability, and magnetoelectric coupling tensors, revealing a bianisotropic moving-medium-like response. This coupling underlies the large synthetic Fizeau-drag effect supported by the wire medium in the homogenization regime.

To illustrate the application of the developed analytical approach, we studied the scattering of electromagnetic waves by a finite thickness space-time-modulated wire medium slab. The reflection and transmission coefficients demonstrate a bianisotropic and non-reciprocal response for obliquely incident plane waves. Finally, we studied the reflection and transmission Goos-Hänchen-like shifts, and showed that for sufficiently long wires these shifts are governed by the Fresnel drag induced by the synthetic velocity. In particular, both shifts can be controlled by the modulation velocity. Our approach may serve as a framework for the analysis of other space-time-modulated materials that admit an analytical treatment based on Lorentz transformations and a co-moving frame.

**Acknowledgements**

This work is supported in part by the Simons Foundation (award SFI-MPS-EWP-00008530-10), by national funds through FCT – Fundação para a Ciência e a Tecnologia, I.P., and, when eligible, co-funded by EU funds under project/support UID/50008/2025 – Instituto de Telecomunicações, with DOI identifier https://doi.org/10.54499/UID/50008/2025.

## Appendix A: Effective material matrix for space-time-modulated wire media

Here, we present an alternative to the Landau-Lifshitz formulation described in Section II.B, in which the effective response in the laboratory frame is expressed directly in terms of the co-moving frame material matrix using an inverse Lorentz transformation.

The effective $6\times 6$ material matrix $\mathbf{M}_{ef}$ formed by permittivity, permeability, and magnetoelectric coupling tensors for moving wire media in the laboratory frame is related to the corresponding response in the co-moving frame by [13]:

$$\mathbf{M}_{ef} = \begin{pmatrix} \varepsilon_0 \bar{\varepsilon}_{ef} & \frac{1}{c}\bar{\xi}_{ef} \\ \frac{1}{c}\bar{\zeta}_{ef} & \mu_0 \bar{\mu}_{ef} \end{pmatrix} = \left[ -\frac{1}{c^2}\mathbf{V} + \mathbf{A}\cdot\mathbf{M}'_{ef} \right] \cdot \left[ \mathbf{A} - \mathbf{V}\cdot\mathbf{M}'_{ef} \right]^{-1}. \tag{A1}$$

Here, $\mathbf{M}'_{ef}$ represents the material matrix in the co-moving frame, which for wire media with strips oriented along the $z-$direction takes the form,

$$\mathbf{M}'_{ef} = \begin{pmatrix} \varepsilon_0 \left( \mathbf{I}_{2\times 2} + \varepsilon'_{zz} \hat{\mathbf{z}} \otimes \hat{\mathbf{z}} \right) & \mathbf{0} \\ \mathbf{0} & \mu_0 \mathbf{I}_{3\times 3} \end{pmatrix}, \tag{A2}$$

where $\varepsilon'_{zz}\left(\omega', k_z\right) = 1 - \frac{\omega_p'^2}{\omega'^2 - c^2 k_z^2}$ and $\mathbf{I}_{2\times 2} = \hat{\mathbf{x}}\otimes\hat{\mathbf{x}} + \hat{\mathbf{y}}\otimes\hat{\mathbf{y}}$, $\mathbf{I}_{3\times 3} = \hat{\mathbf{x}}\otimes\hat{\mathbf{x}} + \hat{\mathbf{y}}\otimes\hat{\mathbf{y}} + \hat{\mathbf{z}}\otimes\hat{\mathbf{z}}$, and $\otimes$ represents a dyadic product of two vectors. Furthermore, with the motion along the $x-$direction we have

$$\mathbf{V} = \begin{pmatrix} \mathbf{0} & \gamma v \left( \hat{\mathbf{z}}\otimes\hat{\mathbf{y}} - \hat{\mathbf{y}}\otimes\hat{\mathbf{z}} \right) \\ -\gamma v \left( \hat{\mathbf{z}}\otimes\hat{\mathbf{y}} - \hat{\mathbf{y}}\otimes\hat{\mathbf{z}} \right) & \mathbf{0} \end{pmatrix}, \tag{A3}$$

$$\mathbf{A} = \begin{pmatrix} \gamma\left( \hat{\mathbf{y}}\otimes\hat{\mathbf{y}} + \hat{\mathbf{z}}\otimes\hat{\mathbf{z}} \right) + \hat{\mathbf{x}}\otimes\hat{\mathbf{x}} & \mathbf{0} \\ \mathbf{0} & \gamma\left( \hat{\mathbf{y}}\otimes\hat{\mathbf{y}} + \hat{\mathbf{z}}\otimes\hat{\mathbf{z}} \right) + \hat{\mathbf{x}}\otimes\hat{\mathbf{x}} \end{pmatrix}. \tag{A4}$$

Substituting Eqs. (A2)-(A4) into Eq. (A1) results in the representation of tensors of effective permittivity, permeability, and magnetoelectric coupling given in Eq. (A1),

$$\bar{\varepsilon}_{ef} = \begin{pmatrix} 1 & 0 & 0 \\ 0 & 1 & 0 \\ 0 & 0 & \dfrac{\varepsilon'_{zz}}{\gamma^2\left(1-\dfrac{v^2}{c^2}\varepsilon'_{zz}\right)} \end{pmatrix}, \tag{A5}$$

$$\bar{\xi}_{ef} = \begin{pmatrix} 0 & 0 & 0 \\ 0 & 0 & 0 \\ 0 & \dfrac{v}{c}\dfrac{\left(\varepsilon'_{zz}-1\right)}{\left(1-\dfrac{v^2}{c^2}\varepsilon'_{zz}\right)} & 0 \end{pmatrix}, \tag{A6}$$

$$\bar{\zeta}_{ef} = \begin{pmatrix} 0 & 0 & 0 \\ 0 & 0 & \dfrac{v}{c}\dfrac{\left(\varepsilon'_{zz}-1\right)}{\left(1-\dfrac{v^2}{c^2}\varepsilon'_{zz}\right)} \\ 0 & 0 & 0 \end{pmatrix}, \tag{A7}$$

$$\bar{\mu}_{ef} = \begin{pmatrix} 1 & 0 & 0 \\ 0 & \dfrac{1}{\gamma^2\left(1-\dfrac{v^2}{c^2}\varepsilon'_{zz}\right)} & 0 \\ 0 & 0 & 1 \end{pmatrix}, \tag{A8}$$

Using the Doppler transformation, $\omega' = \gamma(\omega - v k_x)$ [Eq. (43)], we may write $\varepsilon'_{zz}$ in the above formulas as,

$$\varepsilon'_{zz} = 1 - \frac{\omega_p'^2}{c^2}\frac{1}{\frac{1}{c^2}\gamma^2\left(\omega - v k_x\right)^2 - k_z^2} \tag{A9}$$

the constitutive effective parameters in Eqs. (A5)-(A8) can be obtained as follows:

$$\varepsilon_{ef,zz} = 1 - \frac{\omega_p'^2}{c^2} \frac{\gamma^2}{\frac{\gamma^2}{c^2}\left[\left(\omega - v k_x\right)^2 + \frac{v^2}{c^2}\omega_p'^2\right] - k_z^2}, \tag{A10a}$$

$$\xi_{ef,zy} = \varsigma_{ef,yz} = -\frac{v}{c}\frac{\omega_p'^2}{c^2} \frac{\gamma^2}{\frac{\gamma^2}{c^2}\left[\left(\omega - v k_x\right)^2 + \frac{v^2}{c^2}\omega_p'^2\right] - k_z^2}, \tag{A10b}$$

$$\mu_{ef,yy} = 1 - \frac{v^2}{c^2}\frac{\omega_p'^2}{c^2} \frac{\gamma^2}{\frac{\gamma^2}{c^2}\left[\left(\omega - v k_x\right)^2 + \frac{v^2}{c^2}\omega_p'^2\right] - k_z^2}. \tag{A10c}$$

Interestingly, the magnetoelectric tensors do not exhibit the conventional antisymmetric structure characteristic of isotropic moving media. This difference arises because anisotropic moving media are described by magnetoelectric tensors with a more general structure.

Next, we verify that the material parameters derived above yield the same Landau-Lifshitz effective permittivity tensor as presented in the main text [Eqs. (35a)-(35d)]. This is done with the use of Eq. (19) in [69]:

$$\frac{\overline{\varepsilon}\left(\omega,\mathbf{k}\right)}{\varepsilon_0} = \overline{\varepsilon}_{ef} - \overline{\xi}_{ef}\cdot\overline{\mu}_{ef}^{-1}\cdot\overline{\varsigma}_{ef} + \left(\overline{\xi}_{ef}\cdot\overline{\mu}_{ef}^{-1}\times\frac{\mathbf{k}}{k_0} - \frac{\mathbf{k}}{k_0}\times\overline{\mu}_{ef}^{-1}\cdot\overline{\varsigma}_{ef}\right) + \frac{\mathbf{k}}{k_0}\times\left(\overline{\mu}_{ef}^{-1} - \mathbf{I}_{3\times3}\right)\times\frac{\mathbf{k}}{k_0}, \tag{A11}$$

where $\mathbf{k} = \left(k_x, k_y, k_z\right)$. Substituting Eqs. (A5)-(A8) into Eq. (A11) with $\overline{\mu}_{ef}^{-1} = \hat{\mathbf{x}}\otimes\hat{\mathbf{x}} + \hat{\mathbf{z}}\otimes\hat{\mathbf{z}} + \hat{\mathbf{y}}\otimes\hat{\mathbf{y}}\gamma^2\left(1 - \frac{v^2}{c^2}\varepsilon_{zz}'\right)$, and taking into account Eq. (A9) we obtain after simplifications the following expressions for components of the effective permittivity tensor:

$$\frac{\varepsilon_{xx}}{\varepsilon_0} = 1 - \frac{\omega_p'^2}{c^2}\frac{1}{\frac{1}{c^2}\left[\gamma\left(\omega - k_x v\right)\right]^2 - k_z^2}\left(v\frac{k_z}{\omega}\right)^2\gamma^2, \tag{A12a}$$

$$\frac{\varepsilon_{yy}}{\varepsilon_0} = 1, \tag{A12b}$$

$$\frac{\varepsilon_{zz}}{\varepsilon_0} = 1 - \frac{\omega_p'^2}{c^2} \frac{1}{\frac{1}{c^2}\left[\gamma\left(\omega - k_x v\right)\right]^2 - k_z^2} \left(1 - \frac{k_x}{\omega} v\right)^2 \gamma^2, \tag{A12c}$$

$$\frac{\varepsilon_{xz}}{\varepsilon_0} = \frac{\varepsilon_{zx}}{\varepsilon_0} = -\frac{\omega_p'^2}{c^2} \frac{1}{\frac{1}{c^2}\left[\gamma\left(\omega - k_x v\right)\right]^2 - k_z^2} \left(1 - \frac{k_x}{\omega} v\right)\left(\frac{k_z}{\omega} v\right)\gamma^2. \tag{A12d}$$

Noting that $\frac{\omega_p'^2}{c^2} = \frac{C'/\varepsilon_0}{A_c'}$, $C'L' = \frac{1}{c^2}$, Eqs. (A12a)-(A12d) lead to Eqs. (35a)-(35d), as we wanted to show.

## Appendix B: Reflection and transmission coefficients for a wire medium slab in the co-moving frame

Here, we summarize the expressions for the reflection and transmission coefficients for the electric field for a wire medium slab of thickness $d$ in the static (co-moving) frame. The formulas were originally obtained in [32] for the magnetic field, based on a modal expansion in the air and wire medium slab regions subject to classical and additional boundary conditions at the interfaces.

For TM-polarized obliquely incident plane wave we have:

$$\begin{aligned} R_{\mathrm{TM,co}}(\omega', k_x', k_y') = 1 &- \frac{1}{1 + \frac{\gamma_{\mathrm{TM}}'\left(k_x'^2 + k_y'^2\right)}{\gamma_{z0}'\left(k_x'^2 + k_y'^2 + k_p'^2\right)} \tanh\left(\frac{\gamma_{\mathrm{TM}}' d}{2}\right) - \frac{k_0' k_p'^2}{\gamma_{z0}'\left(k_x'^2 + k_y'^2 + k_p'^2\right)} \tan\left(\frac{k_0' d}{2}\right)} \\ &- \frac{1}{1 + \frac{\gamma_{\mathrm{TM}}'\left(k_x'^2 + k_y'^2\right)}{\gamma_{z0}'\left(k_x'^2 + k_y'^2 + k_p'^2\right)} \coth\left(\frac{\gamma_{\mathrm{TM}}' d}{2}\right) + \frac{k_0' k_p'^2}{\gamma_{z0}'\left(k_x'^2 + k_y'^2 + k_p'^2\right)} \cot\left(\frac{k_0' d}{2}\right)}, \end{aligned} \tag{B1}$$

$$T_{\mathrm{TM,co}}(\omega',k_x',k_y')=\frac{1}{1+\dfrac{\gamma_{\mathrm{TM}}'\left(k_x'^2+k_y'^2\right)}{\gamma_{z0}'\left(k_x'^2+k_y'^2+k_p'^2\right)}\tanh\left(\dfrac{\gamma_{\mathrm{TM}}'d}{2}\right)-\dfrac{k_0'k_p'^2}{\gamma_{z0}'\left(k_x'^2+k_y'^2+k_p'^2\right)}\tan\left(\dfrac{k_0'd}{2}\right)}-\frac{1}{1+\dfrac{\gamma_{\mathrm{TM}}'\left(k_x'^2+k_y'^2\right)}{\gamma_{z0}'\left(k_x'^2+k_y'^2+k_p'^2\right)}\coth\left(\dfrac{\gamma_{\mathrm{TM}}'d}{2}\right)+\dfrac{k_0'k_p'^2}{\gamma_{z0}'\left(k_x'^2+k_y'^2+k_p'^2\right)}\cot\left(\dfrac{k_0'd}{2}\right)}. \tag{B2}$$

Here, $\gamma_{\mathrm{TM}}'=-i\sqrt{k_0'^2-k_x'^2-k_y'^2-k_p'^2}$, $\gamma_{z0}'=-i\sqrt{k_0'^2-k_x'^2-k_y'^2}$, $k_0'=\omega'/c$,

$k_p'^2=(2\pi/p'^2)/[\ln(p'/2\pi r_0')+0.5275]$, $p'=p\gamma$, $r_0'=\gamma w/4$.

For TE-polarized incident plane waves, we have $R_{\mathrm{TE,co}}(\omega',k_x',k_y')=0$ and $T_{\mathrm{TE,co}}(\omega',k_x',k_y')=e^{ik_{z0}'d}$ with $k_{z0}'=\sqrt{k_0'^2-k_x'^2-k_y'^2}$. This result follows from the assumption that the host medium of the wire medium is air. Note that we adopt the $e^{-i\omega't}$ time-variation. In particular, when $\gamma_{\mathrm{TM}}'$ or $\gamma_{z0}'$ are pure imaginary the branch cut of the square root needs to be chosen such that $\mathrm{Im}\{\gamma\}<0$. In addition, the square root branch cut must ensure that $\mathrm{Re}\{\gamma\}\geq 0$.

Furthermore, below we present the expressions for TM and TE reflection coefficients for a wire medium slab of thickness *d* backed by a ground plane in the static (co-moving) frame [50]:

$$R_{\mathrm{TM,co}}(\omega',k_x',k_y')=1-\frac{2}{1+\dfrac{\gamma_{\mathrm{TM}}'\left(k_x'^2+k_y'^2\right)}{\gamma_{z0}'\left(k_x'^2+k_y'^2+k_p'^2\right)}\tanh\left(\gamma_{\mathrm{TM}}'d\right)-\dfrac{k_0'k_p'^2}{\gamma_{z0}'\left(k_x'^2+k_y'^2+k_p'^2\right)}\tan\left(k_0'd\right)}, \tag{B3}$$

$$R_{\mathrm{TE,co}}=-e^{i2k_{z0}'d}. \tag{B4}$$

In particular, Eq. (B3) is used in the calculation of the reflection lateral shift (RS) in Section III.A for the conductor-backed space-time wire medium slab in the laboratory frame.

## Appendix C: Energy balance for plane waves

As it was shown in Section II.A, the total Lorentz force [Eq. (27)] vanishes after being integrated over the space-time-modulated wire medium slab. This property implies that there is no net power absorbed by the metamaterial, i.e. $\int dz \langle p_{loss} \rangle = 0$ in Eq. (21). Here, we derive an independent global energy balance constraint formulated in terms of reflection and transmission matrices.

The conservation of energy for incident, reflected, and transmitted plane waves requires that

$$\left|\mathbf{E}^{inc}\right|^2 = \left|\mathbf{E}^{ref}\right|^2 + \left|\mathbf{E}^{tx}\right|^2 \tag{C1}$$

where for plane waves in free space we have

$$\left|\mathbf{E}\right|^2 = \left|\mathbf{E}_t\right|^2 + \left|E_z\right|^2 = \left|\mathbf{E}_t\right|^2 + \left|\frac{\mathbf{k}_t}{k_{z0}} \cdot \mathbf{E}_t\right|^2 \tag{C2}$$

with $k_{z0} = \sqrt{\left(\frac{\omega}{c}\right)^2 - k_x^2 - k_y^2}$ and $\mathbf{E}_t = E_x \hat{\mathbf{x}} + E_y \hat{\mathbf{y}}$.

Hence, Eq. (C2) can be written as follows:

$$\left|\mathbf{E}\right|^2 = \mathbf{E}_t^* \cdot \underbrace{\left(\mathbf{1} + \frac{1}{\left|k_{z0}\right|^2} \mathbf{k}_t^* \otimes \mathbf{k}_t\right)}_{\mathbf{B}} \cdot \mathbf{E}_t . \tag{C3}$$

This leads to the conclusion that the conservation of energy condition in Eq. (C1) can be expressed in terms of reflection and transmission matrices, $\mathbf{R}$ and $\mathbf{T}$, and the matrix $\mathbf{B}$ defined in Eq. (C3):

$$\mathbf{B} = \mathbf{R}^\dagger \cdot \mathbf{B} \cdot \mathbf{R} + \mathbf{T}^\dagger \cdot \mathbf{B} \cdot \mathbf{T}. \tag{C4}$$

In particular, the absorptance can be evaluated as

$$1-\frac{\mathbf{E}^{inc,*}\cdot\left(\mathbf{R}^{\dagger}\cdot\mathbf{B}\cdot\mathbf{R}+\mathbf{T}^{\dagger}\cdot\mathbf{B}\cdot\mathbf{T}\right)\cdot\mathbf{E}^{inc}}{\mathbf{E}^{inc,*}\cdot\mathbf{B}\cdot\mathbf{E}^{inc}}. \tag{C5}$$

Numerical simulations confirm that (C4) is indeed an identity, demonstrating that the total absorptance vanishes, consistent with $\int dz\langle p_{loss}\rangle=0$ in Eq. (21).

It is relevant to note that in general the absorptance maxima and minima are determined by the eigenvalues of the matrix:

$$\mathbf{B}^{-1/2}\cdot\left[\mathbf{R}^{\dagger}\cdot\mathbf{B}\cdot\mathbf{R}+\mathbf{T}^{\dagger}\cdot\mathbf{B}\cdot\mathbf{T}\right]\cdot\mathbf{B}^{-1/2} \tag{C6}$$

where

$$\mathbf{B}^{-1/2}=\hat{\mathbf{k}}_{tn}^{*}\otimes\hat{\mathbf{k}}_{tn}+\frac{1}{\sqrt{1+\left|k_t/k_{z0}\right|^2}}\hat{\mathbf{k}}_{t}^{*}\otimes\hat{\mathbf{k}}_{t}. \tag{C7}$$

In our problem, both eigenvalues vanish.